\documentclass[twocolumn,preprintnumbers,superscriptaddress]{revtex4}
\usepackage{amsmath}
\usepackage{amssymb}
\usepackage{mathrsfs}
\usepackage{graphicx}
\usepackage{bm}
\usepackage{epstopdf}
\usepackage[colorlinks=true,citecolor=blue,linkcolor=blue,urlcolor=blue]{hyperref}

\begin{document}
\title{Multi-component plasmons in monolayer MoS$_2$ with circularly polarized optical pumping}

\author{Y. M. Xiao}\email{yiming.xiao@foxmail.com}
\affiliation{Department of Physics, University of Antwerp, Groenenborgerlaan 171,
B-2020 Antwerpen, Belgium}
\affiliation{Department of Physics and Astronomy and Yunnan Key Laboratory for
Micro/Nano Materials and Technology, Yunnan University, Kunming 650091, China}

\author{W. Xu}\email{wenxu\_issp@aliyun.com}
\affiliation{Department of Physics and Astronomy and Yunnan Key Laboratory for
Micro/Nano Materials and Technology, Yunnan University,
Kunming 650091, China}
\affiliation{Key Laboratory of Materials Physics, Institute of Solid State Physics,
Chinese Academy of Sciences, Hefei 230031, China}

\author{F. M. Peeters}\email{francois.peeters@uantwerpen.be}
\affiliation{Department of Physics, University of Antwerp, Groenenborgerlaan 171,
B-2020 Antwerpen, Belgium}

\author{B. Van Duppen}\email{ben.vanduppen@uantwerpen.be}
\affiliation{Department of Physics, University of Antwerp, Groenenborgerlaan 171,
B-2020 Antwerpen, Belgium}

\date{\today}

\begin{abstract}
By making use of circularly polarized light and electrostatic gating, monolayer molybdenum 
disulfide (ML-MoS$_2$) can form a platform supporting multiple types of charge carriers. They can be discriminated by their spin, valley index or whether they're electrons or holes. We investigate the collective properties of those charge 
carriers and are able to identify new distinct plasmon modes. We analyze the corresponding dispersion relation, lifetime and oscillator strength, and 
calculate the phase relation between the oscillations in the different components of the plasmon 
modes. All platforms in ML-MoS$_2$ support a long-wavelength $\sqrt{q}$ plasmon branch at zero 
Kelvin. In addition to this, for an $n$-component system, $n-1$ distinct plasmon modes appear as 
acoustic modes with linear dispersion in the long-wavelength limit. These modes correspond to 
out-of-phase oscillations in the different Fermion liquids and have, although being damped, a 
relatively long lifetime. Additionally, we also find new distinct modes at large wave vector that are 
stronger damped by intra-band processes.
\end{abstract}
\maketitle

\section{Introduction}

Recently, monolayers of transition metal dichalcogenides MX$_2$ (M=Mo, W, Nb, Ta, Ti,
and X=S, Se, Te) have been fabricated \cite{Mak10,Splendiani10}. Since then, they are 
drawing intense interest due to their intriguing physical properties. A monolayer MX$_2$ 
(ML-MX$_2$) is a trilayer structure in the form of X-M-X with chalcogen atoms (X) in two
hexagonal planes separated by a plane of metal atoms (M) \cite{Wang12}. Transition metal
dichalcogenides (TMDCs) are indirect band gap semiconductors when stacked in multi-layers
but have a direct band gap in their monolayer form \cite{Mak10,Splendiani10,Wang12}. It
has been shown that field-effect transistors (FETs) made from monolayer MoS$_2$ (ML-MoS$_2$)
could have a room temperature on/off ratio of up to $10^8$ with a mobility higher than 200
cm$^2$/(Vs) \cite{Radisavljevic11ACS,Radisavljevic11, Desai16}. With its ultrathin layered
structure and an appreciable direct band gap, ML-MoS$_2$ has great potential applications
in nano-electronics \cite{Radisavljevic11}, optoelectronics \cite{Lopez-Sanchez13,Lee12},
spintronics and valleytronics \cite{Ganatra14,Zeng12N,Mak12N,Cao12,Mak14}.\par

Investigations of the unique light-matter interaction and many-body effects in ML-MX$_2$
such as photoluminance (PL) \cite{Zeng12N,Mak12N}, optical 
conductivity \cite{Li12,XiaoYM16,Krstajic16}, excitons \cite{He14}, and trions \cite{Mak13NM}
have enriched the understanding of their optical properties. In order to have a better 
understanding of ML-MX$_2$ for potential applications, its plasmonic properties are
also important. 

Plasmons are collective excitations of the electron liquid. They play a 
fundamental role in the dynamical response of electron systems and form the basis of research 
into optical metamaterials \cite{Chen06,Ju11}. Since the discovery of atomically thin 
two-dimensional (2D) graphene \cite{Novoselov04}, it was shown that 2D materials intrinsically 
feature plasmons and could form a platform for potential applications in plasmonic
devices \cite{Grigorenko12}. In recent years, graphene plasmons, in particular, have attracted
a lot of interest because of their unique tunability \cite{Ju11}, long plasmon
lifetime \cite{Koppens11}, and high degree of electromagnetic confinement \cite{Fei12}. This
enables the use of graphene-based plasmonic devices in the spectral range from mid-infrared
to terahertz (THz) \cite{Low14}. The dielectric function of graphene and gapped graphene have 
also been studied intensively and the corresponding polarization functions were obtained 
analytically \cite{Wunsch06,Hwang07,Pyatkovskiy09,Iurov16}. Also plasmons in silicene have been 
investigated with and without external fields \cite{VanDuppen14,Tabert14}. Collective excitations 
of the electron liquid in ML-MoS$_2$ in the absence of external fields have been examined and 
discussed before \cite{Scholz13,Kechedzhi14,Hatami14}.

Due to its band structure, massless Dirac fermions, massive Dirac fermions and a two-dimensional
electron gas can, in principle, all generate a different collective response induced by the inter-particle
Coulomb interaction \cite{Wunsch06,Hwang07,Pyatkovskiy09,Stern67}. In ML-MoS$_2$ charge carriers are described as massive Dirac fermions
(MDF) with a strong intrinsic spin-orbit coupling (SOC). This means that depending on the frequency scale at which one interacts with the material, the response can be similar to a two-dimensional electron gas or to graphene, that has massless Dirac Fermions. Furthermore, the SOC gives rise to a splitting of
conduction and valence bands with opposite spins \cite{Xiao12,Lu13} while preserving the out-of-plane component of the spin as a good quantum number. Moreover, due
to the large band gap in the MDF of ML-MoS$_2$, its low-energy band structure can also be
described by a two-dimensional parabolic band (2DPB) for both the conduction and valence bands
with different valley and spin indices \cite{Kormanyos14, Kormanyos15}. This feature makes the plasmon
dispersion of ML-MoS$_2$ for low frequency plasmons fundamentally different from that of graphene \cite{Wunsch06,Hwang07,Scholz13}.
Indeed, the plasmon dispersion in ML-MoS$_2$ $\omega(q)\sim q^{1/2} n^{1/2}$ is similar to
a traditional two dimensional electron gas (2DEG) while the plasmon dispersion of graphene
is $\omega(q)\sim q^{1/2} n^{1/4}$, where $n$ is the carrier concentration \cite{Scholz13}. For high-frequency plasmons, on the other hand, the relation mimics that of graphene high-frequency plasmons. 

The optical response of ML-MoS$_2$ is governed by the dynamics of charge carriers near the
Dirac points in reciprocal space, i.e. those residing in one of the two valleys of the energy
spectrum. Recently, it was shown that ML-MoS$_2$ has a remarkable valley selective absorption
of circularly polarized light \cite{Zeng12N,Mak12N,Cao12,Xiao12}. This allows one to address
electrons in a single valley. It can be utilized to realize a valley Hall effect \cite{Mak14}. 
As a consequence, the carrier density in ML-MoS$_2$ at a specific valley can be tuned through 
optical pumping \cite{Mak14}.

\begin{figure}[t]
\includegraphics[width=8cm]{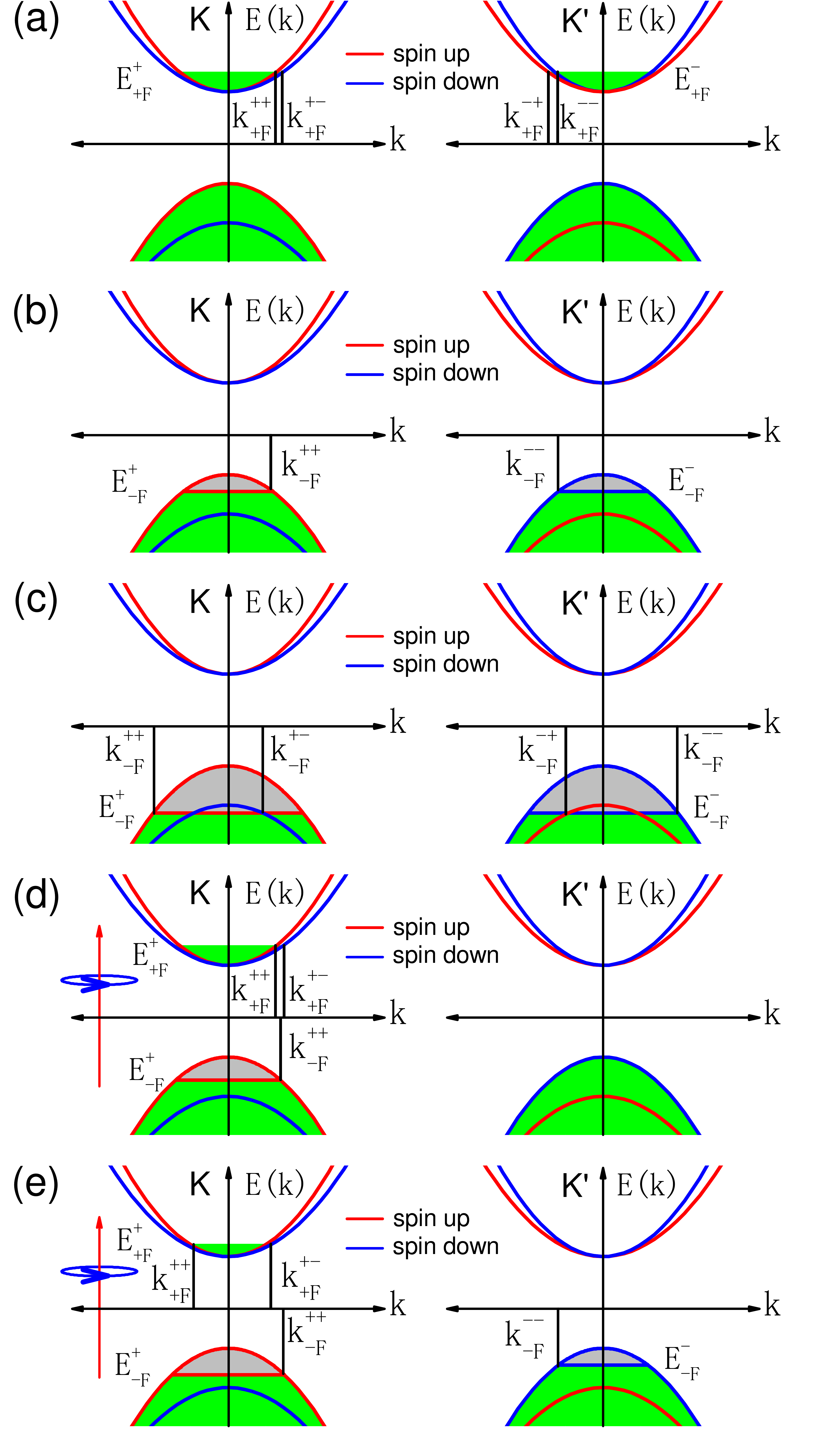}
\caption{(Color online) Band structure in 
the $K$-valley (left column) and $K^{\prime}$-valley (right column) of the different ML-MoS$_2$ systems considered in this study. The bands are colored according 
to the spin type they represent. The green shading indicates electron occupation while the grey 
shading indicates hole occupation. In panels (a) - (c) only electrostatic gating is used to change 
the Fermi level of the systems. In panels (d) - (e) also right-handed circularly polarized light is 
added to the system in order to pump electrons in the $K$-valley from the valence to the conduction 
band. The Fermi vectors $k^{\varsigma s}_{\lambda \rm F}$ and Fermi energies $E_{\rm \lambda F}$ 
indicated follow Eqs. \eqref{Fermieqpp}, \eqref{Fermivec}-\eqref{Fermiener} and 
\eqref{definition_Fermi_vector_energy type1}-\eqref{definition_Fermi_vector_energy}.}
\label{fig1}
\end{figure}

In this way, an optical pumping process can make the electronic system of ML-MoS$_2$ a tunable
multi-component system as shown in Fig. \ref{fig1}. Indeed, by pumping electrons in one
valley from the valence band into the conduction band, one generates a system consisting of
two liquids of interacting Fermions as shown in Figs. \ref{fig1}(d)-(e). By gating the system,
one can change the Fermi level and further fill or empty the conduction or valence bands in
both valleys, allowing for an additional component to appear as shown in Figs. \ref{fig1}(a)
and (c). As the spin-bands are split in ML-MoS$_2$, an additional degree of freedom surfaces
because one can use the Fermi level to access only one of the two spin types per valley in
the valence band, or both of them, as shown in Figs. \ref{fig1}(b)-(c). In this paper, we
investigate the plasmonic response of these different multi-component systems, identify under
which conditions new plasmon modes surface and characterize their properties.

The multi-component Fermion system in ML-MoS$_2$ proves to be a platform for the generation
of a variety of collective effects. Apart from the usual plasmon mode intrinsically present in
a two-dimensional interacting Fermion liquid \cite{Stern67,Giuliani05}, we find that in the
long-wavelength limit an $n$-component system supports $n-1$ lightly damped acoustic modes. 
These modes correspond to oscillations where the different components oscillate with an opposite 
phase. Furthermore, for large wave vectors, we also find new plasmon modes in spectral regions 
where Landau damping occurs for some of the components which then exhibit a high decay rate.

The present paper is organized as follows. Sec. \ref{sec:theoretical approach}, outlines the theory
used to describe plasmons in the multi-component system. In Sec. \ref{sec:theoretical approach A}
we describe the effective low energy band structure of ML-MoS$_2$ with both the MDF and 2DPB models
and point out their differences and similarities. In Sec. \ref{sec:theoretical approach B}, we
evaluate the valley-dependent absorption and calculate the carrier density of a photo-excited
system under circularly polarized optical pumping. In Sec. \ref{sec:theoretical approach C}, we
outline the calculations of the finite-temperature polarization function and how plasmons of
multi-component systems can be calculated within the random phase approximation. We report and discuss the numerical 
results for different multi-component systems in Sec. \ref{sec:results}. The optical absorption of 
circularly polarized light in ML-MoS$_2$ are presented in Sec. \ref{sec:results} A. We compare
the results of massive and hyperbolic models for $n$-type ML-MoS$_2$ in Sec. \ref{sec:results} B.
In Sec. \ref{sec:results} C, a spin-polarized two-component system at finite temperature is
discussed. The valley-polarized three- and four-component systems are discussed in
Sec. \ref{sec:results} D and E, respectively. Finally, our main conclusions are summarized in Sec. \ref{sec:conclusions}.

\section{Theoretical approach}
\label{sec:theoretical approach}

\subsection{Electronic band structure and carrier concentration}
\label{sec:theoretical approach A}
In this study, we consider a ML-MoS$_2$ sheet positioned in the x-y plane. The effective
Hamiltonian for a carrier (an electron or a hole)  around the $K$ and $K^{\prime}$  points in 
reciprocal space can be written as \cite{Xiao12,Lu13}
\begin{equation}
\hat{H}^\varsigma_0=\left(
\begin{array}{cc}
\Delta/2 &\varsigma atk_{-\varsigma}  \\
\varsigma atk_\varsigma &-\Delta/2+\varsigma s\gamma \\
\end{array}
\right),
\end{equation}
where $\mathbf{k}=(k_x,k_y)$ being the wavevector, $k_{\pm}=k_x\pm ik_y=ke^{\pm i\theta}$, and
$\theta$ is the angle between $\mathbf{k}$ and the $k_{x}$-axis, $\varsigma$ is the valley index
with $\varsigma = +$ for $K$ and $\varsigma = -$ for the $K^{\prime}$  valley, the spin index $s=\pm$ for
spin-up and spin-down states, the lattice parameter $a=3.193$ {\AA}, the hopping parameter
$t=1.1$ eV \cite{Xiao12}, the spin-orbit parameter $\gamma=75$ meV \cite{Zhu11}, and $\Delta$
is the direct band gap equal to 1.66 eV between the conduction and valence bands \cite{Xiao12,Lu13}. Notice that DFT calculations \cite{Kormanyos15} have attributed a slightly larger mass to the holes than to electrons in ML-MoS$_2$. This has not been accounted for in the model described above but is not expected to have a significant impact on the results presented. \par

The corresponding Schr\"{o}dinger equation for ML-MoS$_2$ at valley $K$ or $K^{\prime}$  can be solved
analytically and the eigenvalues are given by
\begin{equation}\label{Dispersion}
E^{\varsigma s}_{\lambda\mathbf{k}}=\varsigma s\gamma/2
+\lambda\Lambda^{\varsigma s}(\mathbf{k}),
\end{equation}
where $\Lambda^{\varsigma s}(\mathbf{k})=[a^2t^2\mathbf{k}^2+\Delta_{\varsigma s}^2]^{1/2}$,
$\Delta_{\varsigma s}=(\Delta-\varsigma s\gamma)/2$, and $\lambda=\pm$ refers to conduction/valence
band. The corresponding eigenfunction for a carrier in conduction or valence bands near the
$K$($K'$) point denoted by $|\mathbf{k}, \lambda\varsigma s\rangle$ can be written as
\begin{align}
&|\mathbf{k}, +\varsigma s\rangle=[\cos(\vartheta_\mathbf{k}^{\varsigma s}/2), \varsigma\sin(\vartheta_\mathbf{k}^{\varsigma s}/2)e^{i\varsigma\theta}]
e^{i\mathbf{k}\cdot\mathbf{r}},\nonumber\\
&|\mathbf{k}, -\varsigma s\rangle=[-\sin(\vartheta_\mathbf{k}^{\varsigma s}/2), \varsigma\cos(\vartheta_\mathbf{k}^{\varsigma s}/2)e^{i\varsigma\theta}]
e^{i\mathbf{k}\cdot\mathbf{r}},
\end{align}
respectively, in a form of row vector with
\begin{align}
\cos\vartheta^{\varsigma s}_\mathbf{k}=\frac{\Delta_{\varsigma s}}{\sqrt{a^2t^2k^2+\Delta^2_{\varsigma s}}},
\sin\vartheta^{\varsigma s}_\mathbf{k}=\frac{atk}{\sqrt{a^2t^2k^2+\Delta^2_{\varsigma s}}}.\nonumber
\end{align}

The electron (hole) density $n^\varsigma_\lambda$ ($\lambda=+$ for conduction band and $\lambda=-$
for valence band) of ML-MoS$_2$ at valley $\varsigma$ can be written as
\begin{equation}\label{Cdensi}
n^\varsigma_{\lambda}=\frac{1}{(2\pi)^2}\sum_{s=\pm}\int d\mathbf{k}^2
[\delta_{\lambda,-1}+\lambda f_\lambda(E^{\varsigma s}_{\lambda\mathbf{k}})],
\end{equation}
where $f_\lambda(E^{\varsigma s}_{\lambda\mathbf{k}})=[e^{(E^{\varsigma s}_{\lambda\mathbf{k}}-\mu^\varsigma_\lambda)/k_B T}+1]^{-1}$ is the Fermi-Dirac distribution
function for electrons and $\mu^\varsigma_\lambda$ is the chemical potential (or Fermi energy $E^\varsigma_{\lambda\mathrm{F}}$ at zero temperature) for electrons in conduction band or
holes in valence band at $\varsigma$ valley for a photo-excited quasi-equilibrium system in
the present study. At $T$=0 K, Eq. (\ref{Cdensi}) reduces to the familiar relation between the
carrier density and the Fermi vector $k^{\varsigma s}_{\lambda\mathrm{F}}$ for a specific
conduction/valence subband with spin index $s$ and valley index $\varsigma$,
\begin{equation}\label{CdeS}
n^\varsigma_{\lambda}=\sum_{s=\pm}[k^{\varsigma s}_{\lambda\mathrm{F}}]^2/(4\pi).
\end{equation}

As depicted in Fig. \ref{fig1}, one can use the Fermi level to generate charge carrier liquids
with different spins. If the Fermi level lies above the top point of the lowest valence subband for
a $p$-type ML-MoS$_2$ as shown in Fig. \ref{fig1} (b), the lowest subband is fully occupied by
electrons and the holes are only distributed in the upper valence subband. The hole density
in the upper valence subband is denoted by $n^\varsigma_-$. This regime holds when the Fermi
level satisfies $E^{\varsigma}_{-\mathrm{F}}\geq -\Delta/2-\gamma$, which corresponds to a hole
density $n^\varsigma_-\leq (\gamma^2+\Delta\gamma)/(2\pi a^2t^2)=1.679\times 10^{13}$ cm$^{-2}$
at $\varsigma$ valley. The Fermi wave vector and Fermi level for the upper valence subband at
valley $\varsigma$ are given by
\begin{align}\label{Fermieqpp}
&[k^{\varsigma s}_{-\mathrm{F}}]^2=4\pi n^\varsigma_{-}(\varsigma s=1),\nonumber\\
&E^{\varsigma}_{-\mathrm{F}}=\gamma/2-[4\pi a^2t^2n^\varsigma_{-}
+(\Delta-\gamma)^2/4]^{1/2}.
\end{align}

Then, we consider the other situations as shown in Fig. \ref{fig1}(a) and Fig. \ref{fig1}(c). 
Using the relation that the Fermi energies for Fermi wavevectors of spin-up and spin-down 
subbands should be equal, we obtain
\begin{equation}\label{Fermieq}
\Lambda^{\varsigma-}(k^{\varsigma-}_{\lambda \mathrm{F}})-
\Lambda^{\varsigma+}(k^{\varsigma+}_{\lambda \mathrm{F}})=\varsigma\lambda\gamma.
\end{equation}
After combining Eqs. \eqref{CdeS} and \eqref{Fermieq}, one can derive the roots of
$[k^{\varsigma s}_{\lambda \mathrm{F}}]^2$ as
\begin{align}\label{Fermivec}
[k^{\varsigma s}_{\lambda \mathrm{F}}]^2=2\pi n^\varsigma_\lambda
+\varsigma s\bigg[\frac{\Delta\gamma}{2a^2t^2}
-\lambda\frac{[\Delta^2\gamma^2
+8\pi n^\varsigma_\lambda a^2t^2\gamma^2]^{1/2}}{2a^2t^2}\bigg].
\end{align}
Thus, the Fermi energy for a zero temperature system at valley $\varsigma$ with a carrier
density $n^\varsigma_{\lambda}$ is given by
\begin{equation}\label{Fermiener}
E^{\varsigma}_{\lambda\mathrm{F}}=\varsigma s\gamma/2
+\lambda\Lambda^{\varsigma s}(k^{\varsigma s}_{\lambda \mathrm{F}}).
\end{equation}
The carrier density in a specific electronic branch can also be written as
\begin{align}
n_{\lambda}^{\varsigma s}=\frac{n^\varsigma_\lambda}{2}+\varsigma s
\bigg[\frac{\Delta\gamma}{8\pi a^2t^2}-\frac{\lambda[\Delta^2\gamma^2
+8\pi n^\varsigma_\lambda\gamma^2a^2t^2]^{1/2}}{8\pi a^2t^2}\bigg].
\end{align}

The density of state (DOS) can be presented by the imaginary part of the Green function as
\begin{align}
D_\lambda(E)=\sum_{\varsigma s}\frac{|E|-\lambda\varsigma s\gamma/2}
{2\pi a^2t^2}\theta[|E|-\lambda\varsigma s\gamma/2-\Delta_{\varsigma s}].
\end{align}

When the carrier concentration is low, one can expand Eq. (\ref{Dispersion}) for small deviations 
from the $K$ or $K^{\prime}$ point. In that case, the low energy electronic band structure of 
ML-MoS$_2$ can be written as the 2DPB form for free electron/hole gases as
\begin{equation}\label{2Dproxi}
\tilde{E}^{\varsigma s}_{\lambda\mathbf{k}}=
\lambda\frac{a^2t^2\mathbf{k}^2}{2\Delta_{\varsigma s}}
+\lambda\Delta_{\varsigma s}+\frac{\varsigma s\gamma}{2},
\end{equation}
and the DOS is given by
\begin{align}
\tilde{D}_\lambda(E)
=\sum_{\varsigma s}\frac{\Delta_{\varsigma s}}{2\pi a^2t^2}
\theta[|E|-\lambda\varsigma s\gamma/2-\Delta_{\varsigma s}].
\end{align}
For a $p$-type sample at valley $\varsigma$ with a hole density
$n_-^\varsigma\leq(\Delta-\gamma)\gamma/(2\pi a^2t^2)$, the Fermi vector and Fermi level for
the upper valence band is
\begin{align}\label{definition_Fermi_vector_energy type1}
&[\tilde{k}^{\varsigma s}_{- \mathrm{F}}]^2=4\pi
n_\lambda^\varsigma (\varsigma s=1),\nonumber\\
&\tilde{E}^{\varsigma}_{-\mathrm{F}}=-4\pi a^2t^2
n_\lambda^\varsigma/(\Delta-\gamma)-\Delta/2+\gamma.
\end{align}
For the other cases with $n$- or $p$-type doping, the Fermi vector and Fermi level for a specific
spin and valley subband can be written as
\begin{align}
&[\tilde{k}^{\varsigma s}_{\lambda \mathrm{F}}]^2=2\pi n_\lambda^\varsigma-\varsigma s\bigg[\frac{2\pi n_\lambda^\varsigma\gamma}{\Delta}-\frac{(\Delta^2-\gamma^2)\gamma}{a^2t^2\Delta}\delta_{\lambda,-1}\bigg],\nonumber\\
&\tilde{E}^{\varsigma}_{\lambda\mathrm{F}}=\lambda\frac{a^2t^2[\tilde{k}^{\varsigma s}_{\lambda \mathrm{F}}]^2}
{2\Delta_{\varsigma s}}+\lambda\Delta_{\varsigma s}+\frac{\varsigma s\gamma}{2}.
\label{definition_Fermi_vector_energy}
\end{align}

\subsection{Quasi-equilibrium system by optical pumping}
\label{sec:theoretical approach B}

In this section, we solve the Boltzmann equation to obtain the response of charge carriers
with different spin- and valley indexes to circularly polarized light. This enables us to
find the quasi-equilibrium electron and hole densities in each valley that will be used in
the next section to investigate the plasmonic response.

Within the Coulomb gauge, the vector potential of a light field with
left ($\nu=-$)/right ($\nu=+$) handed polarization is given by \cite{Ibanez-Azpiroz12}
\begin{align}
A_{\nu}(t)=\frac{F_0}{\sqrt{2}\omega}\sin(\omega t)
(\hat{\mathbf{x}}+\nu i\hat{\mathbf{y}}).
\end{align}
Within first-order perturbation theory, the steady-state electronic transition rate of
inter-band transitions between valence and conduction band induced by direct
carrier-photon interaction can be obtained by using the Fermi's golden rule, which reads
\begin{align}
W^{\varsigma s,\mp}_{\nu,\lambda\lambda'}(\mathbf{k},\mathbf{k'})=
&\frac{2\pi}{\hbar}\bigg(\frac{eatF_0}{\sqrt{2}\hbar\omega}\bigg)^2
\delta_{\mathbf{k},\mathbf{k'}}\delta(E^{\varsigma s}_{\lambda'\mathbf{k'}}
-E^{\varsigma s}_{\lambda\mathbf{k}}\mp\hbar\omega)\nonumber\\
&\times[\cos^4(\vartheta_\mathbf{k}^{\varsigma s}/2)\delta_{\varsigma\nu,1}
+\sin^4(\vartheta_\mathbf{k}^{\varsigma s}/2)\delta_{\varsigma\nu,-1}],
\end{align}
where $\lambda'=-\lambda$, the Delta function $\delta_{\varsigma\nu,\pm1}$ indicates the
valley-dependent selection rule for optical transitions in ML-MoS$_2$ under a circularly
polarized light field, and the $\mp$ sign in the Delta function refers to absorption ($-$) 
or emission ($+$) of a photon with energy $\hbar\omega$.

Due to the fast spin relaxation, compared to the photo-excited carrier life
time \cite{Mak14,Song13,Korn11}, the photo-excited system in this study can be considered 
as a quasi-equilibrium system where the carrier distributions in conduction and valence 
bands can be approximately described by the Fermi-Dirac function with separate chemical 
potentials for electrons in conduction band and holes in valence band. For nondegenerate 
statistics, the Boltzmann equation (BE) for each valley and spin subsystem takes the form
\begin{align}
\frac{\partial f^{\varsigma s}_\lambda(\mathbf{k})}{\partial t}
=\sum_{\lambda',\mathbf{k'}} F_\nu^{\varsigma s}(\mathbf{k},\mathbf{k'})
-\frac{f^{\varsigma s}_\lambda(\mathbf{k})-f^{\varsigma s}_{\lambda0}
(\mathbf{k})}{\tau},
\end{align}
with $F_\nu^{\varsigma s}(\mathbf{k},\mathbf{k'})=[W^{\varsigma s,-}_{\nu, \lambda'\lambda}
(\mathbf{k},\mathbf{k'})+W^{\varsigma s,+}_{\nu, \lambda'\lambda}
(\mathbf{k},\mathbf{k'})][f^{\varsigma s}_{\lambda'}(\mathbf{k'})
-f^{\varsigma s}_{\lambda}(\mathbf{k})]$, and $f^{\varsigma s}_{\lambda}(\mathbf{k})\simeq
f_{\lambda}(E_{\lambda\mathbf{k}}^{\varsigma s})$ is the carrier momentum distribution
function and $f^{\varsigma s}_{\lambda0}(\mathbf{k})$ represents the initial state in the 
presence of a dark system.

For the first moment, the mass-balance equation (or rate equation) can be derived after
summing both sides of the BE over spin and wave vector. This leads to
\begin{align}
\frac{\partial n^\varsigma_+}{\partial t}=\frac{\partial n^\varsigma_-}{\partial t}
=&\sum_{s}G_\nu^{\varsigma s}-\frac{\Delta n^\varsigma}{\tau},
\end{align}
where the carrier generation rate within each subsystem is
\begin{align}
G^{\varsigma s}_\nu=&\frac{e^2F_0^2}{32\hbar^2\omega}\bigg[1+\nu\varsigma
\frac{2\Delta_{\varsigma s}}{\hbar\omega}\bigg]^2
\theta(\hbar\omega-2\Delta_{\varsigma s})\nonumber\\
&\times\bigg[f_-\bigg(\frac{\varsigma s\gamma-\hbar\omega}{2}\bigg)-
f_+\bigg(\frac{\varsigma s\gamma+\hbar\omega}{2}\bigg)\bigg],
\end{align}
$\Delta n^\varsigma$ is the photo-excited carrier density and $\tau$ is the photo-excited
carrier life time which can be measured experimentally \cite{Korn11}. In the steady quasi-equilibrium 
state of the system, i.e., for $dn^\varsigma_+/dt=dn^\varsigma_-/dt=0$, the mass-balance 
equation at valley $\varsigma$ becomes
\begin{align}\label{photo-excited}
\Delta n^\varsigma=\tau\sum_{s}G_\nu^{\varsigma s}.
\end{align}

When one pumps the system with a circularly polarized optical beam, the electrons in the
valence band are excited into the conduction band such that a gas of photo-excited carriers is formed in the conduction band. Therefore, the chemical potential for electrons/holes in conduction/valence band
for each valley can be determined through
\begin{align}\label{EHden}
n^\varsigma_\lambda&=n^\varsigma_{\lambda0}+\Delta n^\varsigma.
\end{align}
where $n^\varsigma_{\lambda0}$ is the initial carrier density in conduction or valence band.
After combining Eqs. \eqref{photo-excited} and \eqref{EHden}, the photo-excited carrier
density can be determined.

From this, also the optical conductivity can be obtained as
\begin{align}
\sigma^{\varsigma s}_\nu(\omega)=&\frac{2\hbar\omega G^{\varsigma s}_\nu}{F_0^2}
=\sigma_0\bigg[1+\varsigma\nu\frac{2\Delta_{\varsigma s}}{\hbar\omega}\bigg]^2
\theta(\hbar\omega-2\Delta_{\varsigma s})\nonumber\\
&\times\bigg[f_-\bigg(\frac{\varsigma s\gamma-\hbar\omega}{2}\bigg)-
f_+\bigg(\frac{\varsigma s\gamma+\hbar\omega}{2}\bigg)\bigg],
\end{align}
where $\sigma_0=e^2/(16\hbar)$. Finally, we can define the degree of valley-dependent 
absorption (VA) for light with polarization $\nu$ as
\begin{align}
P_\nu(\omega)=\frac{\sum_s[\sigma^{+s}_\nu(\omega)-\sigma^{-s}_\nu(\omega)]}
{\sum_s[\sigma^{+s}_\nu(\omega)+\sigma^{-s}_\nu(\omega)]}.
\end{align}
This quantity describes the difference between the absorption (or photo-excited carrier density)
at $K$ and $K^{\prime}$  valleys under circularly polarized optical pumping.

\subsection{Polarization function and plasmons}
\label{sec:theoretical approach C}

In this section, we set up the theoretical framework to calculate plasmons in a multi-component
system. We work within the Random Phase Approximation (RPA).

When the degeneracy of spin or valley degrees of freedom of a electron/hole gas system is broken, it can
be regarded as a multi-component system. For a multi-component system, the component-resolved
response functions $\Pi^{i,i'}(\mathbf{q},\omega)$ of the interacting electron liquid
are given, in the RPA, by the following matrix equation \cite{Giuliani05,Agarwal14}
\begin{align}\label{multi-matrix}
[\Pi^{i,i'}(\mathbf{q},\omega)]^{-1}=\{\Pi^{i}(\mathbf{q},\omega)\}^{-1}
\delta_{i,i'}-v_q,
\end{align}
where $\Pi^{i}(\mathbf{q},\omega)$ is the non-interacting response function for the $i$-th component
system and $v_q=e^2/(2\epsilon_\mathrm{r}\epsilon_0q)$ is the Fourier transform of the Coulomb
interaction with $\epsilon_{\rm r}=5$ the relative dielectric constant of the background \cite{Scholz13,Kechedzhi14}. The total
density-density response function within the RPA can then be written as
\begin{equation}
\Pi^\mathrm{RPA}(\mathbf{q},\omega)=\sum_{ii'}\Pi^{i,i'}(\mathbf{q},\omega)
=\sum_{i}\Pi^{i}(\mathbf{q},\omega)/\epsilon_\mathrm{RPA}(\mathbf{q},\omega),
\end{equation}
where the RPA dielectric function is defined as
\begin{equation}
\epsilon_\mathrm{RPA}(\mathbf{q},\omega)=1-v_q\sum_i\Pi^{i}(\mathbf{q},\omega).
\end{equation}

Plasmons in the RPA can be found from the zeros of the dielectric function $\epsilon_\mathrm{RPA}(\mathbf{q},\omega(q)-i\eta)$, where $\omega(q)$ is the plasmon
frequency and $\eta$ the decay rate of the plasmon \cite{Pyatkovskiy09}. Usually, plasmons
can be approximately determined by the roots of the real part dielectric function
$ \mathrm{Re}[\epsilon_\mathrm{RPA}(\mathbf{q},\omega)]=0$ \cite{VanDuppen14,Tabert14},
which should also lead to resonance peaks in the energy loss function
$\mathrm{Los}(\mathbf{q},\omega)=-\mathrm{Im}[\epsilon_\mathrm{RPA}]^{-1}$ that can be
measured by means of electron energy loss spectroscopy (EELS) \cite{Politano14}. The decay rate 
(inverse life time) of the weakly damped plasmon is given by
\begin{align}
\eta=\frac{\mathrm{Im}[\Pi(\mathbf{q},\omega(q))]}{\{(\partial/\partial\omega) \mathrm{Re}[\Pi(\mathbf{q},\omega)]\}_{\omega=\omega(q)}}.
\end{align}

The imaginary part of the dynamical RPA polarization near the undamped plasmon branch is given
by \cite{Giuliani05,VanDuppen14}
\begin{align}\label{undamped oci}
\mathrm{Im}[\Pi^{\mathrm{RPA}}(\mathbf{q},\omega(q))]=-O(\omega(q))\delta(\omega-\omega(q)),
\end{align}
and the oscillator strength of the undamped plasmon mode is defined as
\begin{align}
O(\omega(q))=\frac{-\pi}{v_q}\frac{\mathrm{Re} [\Pi(\mathbf{q},\omega(q))]}
{\{(\partial/\partial\omega)\mathrm{Re} [\Pi(\mathbf{q},\omega)]\}_{\omega=\omega(q)}}.
\end{align}
We use the strength of the absorption spectral function
$A(\mathbf{q},\omega)=-\mathrm{Im}[\Pi^{\mathrm{RPA}}(\mathbf{q},\omega)]$ \cite{Kechedzhi14} to
describe the oscillator strength of the damped plasmon modes in the
particle-hole excitation spectrum (PHES) ($\mathrm{Im}[\Pi(\mathbf{q},\omega)]\neq 0$).

For both the undamped and weakly damped plasmon modes, the amplitude of the plasmon oscillation
of each component $N(i)$ can be obtained by calculating the eigenmodes \cite{Agarwal14} of the
real part of the matrix equation \eqref{multi-matrix}
\begin{align}
[\{\mathrm{Re}[\Pi^{i}(\mathbf{q},\omega(q))]\}^{-1}\delta_{i,i'}-v_q]N(i)=0,
\end{align}
from which we can get the ratio of the plasmon oscillation amplitude
\begin{align}\label{Osratio}
N(i)/N(i')=\mathrm{Re}[\Pi^{i}(\mathbf{q},\omega(q))]/\mathrm{Re}[\Pi^{i'}
(\mathbf{q},\omega(q))].
\end{align}

We now turn to the calculation of the non-interacting response functions for each
component $\Pi^{i}$. We denote each component by the spin and valley index, i.e.
$i=\varsigma, s$ and find
\begin{equation}\label{polarization}
\Pi^{\varsigma s}(\mathbf{q},\omega)=\sum_{\lambda\lambda'\mathbf{k}}
\frac{f(E^{\varsigma s}_{\lambda\mathbf{k}})-
f(E^{\varsigma s}_{\lambda'\mathbf{k+q}})}
{\hbar\omega+E^{\varsigma s}_{\lambda\mathbf{k}}
-E^{\varsigma s}_{\lambda'\mathbf{k+q}}+i\delta}
C^{\varsigma s}_{\lambda\mathbf{k},\lambda'\mathbf{k+q}},
\end{equation}
where the structure factor
\begin{align}\label{sfactor}
C^{\varsigma s}_{\lambda\mathbf{k},\lambda'\mathbf{k+q}}
=\frac{1}{2}\bigg[1+\lambda\lambda'\frac{\Delta_{\varsigma s}^2
+a^2t^2\mathbf{k}(\mathbf{k+q})}{\Lambda^{\varsigma s}(\mathbf{k})
\Lambda^{\varsigma s}(\mathbf{k+q})}\bigg],
\end{align}
with $\varphi$ the angle between $\mathbf{k}$ and $\mathbf{k+q}$ with 
$\cos\varphi=(k+q\cos\phi)/|\mathbf{k+q}|$, $\phi$ is the angle between $\mathbf{k}$
and $\mathbf{q}$, and $f(E^{\varsigma s}_{\lambda\mathbf{k}})=f_\lambda(E^{\varsigma s}_{\lambda\mathbf{k}})$ 
is the Fermi-Dirac distribution function for electrons.

At long-wavelength ($\mathbf{q}\rightarrow0$) and low-temperature ($T\rightarrow0$ K), we can
expand Eq. \eqref{sfactor} to second order in $q$:
\begin{align}
C^{\varsigma s}_{\lambda\mathbf{k},\lambda'\mathbf{k+q}}\simeq\delta_{\lambda,\lambda'}-\lambda\lambda'
\sin^2(\vartheta_{\mathbf{k}}^{\varsigma s})q^2\sin^2\phi/(4k^2).
\end{align}
For intra-band transitions ($\lambda'=\lambda$), the Lindhard ratio can be expanded to second order in $q$
with the result
\begin{align}
\frac{f(E^{\varsigma s}_{\lambda\mathbf{k}})-
f(E^{\varsigma s}_{\lambda\mathbf{k+q}})}
{\hbar\omega+E^{\varsigma s}_{\lambda\mathbf{k}}
-E^{\varsigma s}_{\lambda\mathbf{k+q}}}
=&q\cos\phi\delta(E^{\varsigma s}_{\lambda\mathbf{k}}
-E^\varsigma_{\lambda\mathrm{F}})
\frac{\partial E^{\varsigma s}_{\lambda\mathbf{k}}}{\partial k}\nonumber\\
&\times\bigg(\frac{1}{\hbar\omega}+\frac{q\cos\phi}{\hbar^2\omega^2}
\frac{\partial E^{\varsigma s}_{\lambda\mathbf{k}}}{\partial k}\bigg).
\end{align}
Because $C^{\varsigma s}_{\lambda\mathbf{k},\lambda'\mathbf{k+q}}\simeq1$
in the long-wavelength limit, we obtain the real part of the intra-band part of the polarization function as
\begin{align}
\mathrm{Re}[\Pi_\mathrm{intra}(\mathbf{q},\omega)]=&\frac{q^2}{4\pi\hbar^2\omega^2}
\sum_{\lambda\varsigma s}(u^\lambda_{\varsigma s}-\Delta^2_{\varsigma s}/u^\lambda_{\varsigma s})\nonumber\\
&\times\theta(u^\lambda_{\varsigma s}-\Delta_{\varsigma s}),
\end{align}
with $u^\lambda_{\varsigma s}=|E^{\varsigma}_{\lambda\mathrm{F}}|-\lambda\varsigma s\gamma/2$
where $E^{\varsigma}_{\lambda\mathrm{F}}$ is the Fermi energy for electrons ($\lambda=1$) in the
conduction band or holes ($\lambda=-1$) in the valence band.\par

For inter-band transitions ($\lambda'=-\lambda$), the real part of the polarization function can
be expanded to second order in $q$ with
\begin{align}
\mathrm{Re}[\Pi_\mathrm{inter}(\mathbf{q},\omega)]=&\frac{-q^2}{32\pi\hbar\omega}\sum_{\lambda\varsigma s}\ln
\bigg|\frac{\hbar\omega+2u^\lambda_{\varsigma s}}{\hbar\omega-2u^\lambda_{\varsigma s}}\bigg|\nonumber\\
&\times\theta(u^\lambda_{\varsigma s}-\Delta_{\varsigma s}).
\end{align}
This only contributes by a small amount to the total polarization function in the low frequency
regime where plasmons exist.\par

After using the low-$q$ expansion of the polarization function and neglecting the logarithmic
correction, the charge plasmon dispersion in ML-MoS$_2$ is given by
\begin{equation}
\omega_0(q)=\big[e^2q\sum_{\lambda\varsigma s}(u^\lambda_{\varsigma s}-\Delta^2_{\varsigma s}
/u^\lambda_{\varsigma s})\theta(u^\lambda_{\varsigma s}-\Delta_{\varsigma s})/(8\pi\epsilon_\mathrm{r}\epsilon_0)\big]^{1/2}.
\label{square_root_plasmon}
\end{equation}

In order to calculate the polarization at arbitrary wave vector $q$, we note that the individual
valley and spin resolved systems can be regarded as analogous to gapped graphene or
silicene \cite{Pyatkovskiy09,VanDuppen14,Tabert14}. The polarization function depends on
$u^\lambda_{\varsigma s}$ which corresponds to a shift of $\pm\gamma/2$ away from the absolute
value of the Fermi level. At zero temperature, the polarization function of an electron liquid
with spin $s$ in the $\varsigma$ valley can be written as
\begin{align}
\Pi_{T=0}^{\varsigma s}(\mathbf{q},\omega)=&\Pi^{\varsigma s}_{0,T=0}(\mathbf{q},\omega)
[\tilde{\theta}(\Delta_{\varsigma s}-u^-_{\varsigma s})
\tilde{\theta}(\Delta_{\varsigma s}-u^+_{\varsigma s})\nonumber\\
&-\theta(u^-_{\varsigma s}-\Delta_{\varsigma s})
\theta(u^+_{\varsigma s}-\Delta_{\varsigma s})]\nonumber\\
&+\sum_\lambda\Pi^{\varsigma s\lambda}_{1,T=0}(\mathbf{q},\omega)
\theta(u^\lambda_{\varsigma s}-\Delta_{\varsigma s}),
\end{align}
with the step functions
\begin{equation}
\tilde{\theta}(x)=\bigg\{
\begin{array}{c}
1 ,x\geqslant0\\
0 ,x<0\\
\end{array},
\theta(x)=\bigg\{
\begin{array}{c}
1 ,x>0\\
0 ,x\leqslant0\\
\end{array}.
\end{equation}
The detailed polarization function in the ($q$-$\omega$) plane is presented in the Appendix.

Finally, we note that in order to calculate the response of the system at finite temperature,
one can use the following identity to express the Fermi-Dirac function \cite{Maldague78}
\begin{align}
\frac{4k_BT}{e^{(E-\mu)/k_BT}+1}=\int_{-\infty}^\infty\frac{d\mu'
\theta(\mu'-E)}{\cosh^2[(\mu'-\mu)/2k_BT]}.
\end{align}
Thus, the polarization function could be obtained as an integral transformation of its
corresponding zero-temperature polarization function as \cite{Tomadin13}
\begin{align}
\Pi&^{\varsigma s}_{T}(\mathbf{q},\omega; \mu^\varsigma_{+},\mu^\varsigma_{-})
=\int_{\Delta/2}^\infty
\frac{d\mu'\Pi^{\varsigma s+}_{1,T=0}(\mathbf{q},\omega)|_{E^\varsigma_{+ \mathrm{F}}=\mu'}}{4k_BT\cosh^2[(\mu'-\mu^\varsigma_+)/2k_BT]}\nonumber\\
&+\int^{C}_{-\infty}
\frac{d\mu'\Pi^{\varsigma s-}_{1,T=0}(\mathbf{q},\omega)|_{E^\varsigma_{- \mathrm{F}}=\mu'}}{4k_BT\cosh^2[(\mu'-\mu^\varsigma_-)/2k_BT]}
+\Pi^{\varsigma s}_{0,T=0}(\mathbf{q},\omega)\nonumber\\
&\times[\mathcal{F}_T(C-\mu^\varsigma_-)-
\mathcal{F}_T(\Delta/2-\mu^\varsigma_+)],
\end{align}
where $C=\varsigma s\gamma-\Delta/2$ and $\mathcal{F}_T(x)=(e^{x/k_BT}+1)^{-1}$,
and $\mu^{\varsigma}_{\lambda}$ is the chemical potential for conduction
and valence bands at valley $\varsigma$ which can be obtained through Eq. \eqref{Cdensi}.

If the ML-MoS$_2$ system has low carrier density, one can also use a two dimensional
parabolic band (2DPB) model to describe the electronic structure as explained in the previous
section. The polarization function and plasmons of ML-MoS$_2$ with the 2DPB model are
presented in the Appendix.

\section{Results and discussions}
\label{sec:results}

In this study, we consider both $n$- and $p$-doped ML-MoS$_2$, in the presence and absence
of circularly polarized light. We can safely ignore the Rashba spin-orbit coupling because it
requires a strong external perpendicular electric field \cite{XiaoYM16}. Therefore, the optical
field only couples to the orbital part of the wave function and the spin is conserved during
optical transitions. It should be noted that the determination of the photo-excited density in
the presence of circularly polarized light in Sec. \ref{sec:theoretical approach B} is only
suitable for a relatively weak light fields. For high carrier density generated by optical pumping, 
we extracted the photo-excited carrier density from experimental data~\cite{Mak14}. In the following 
sections, we present numerical calculations for both the plasmon dispersion, energy loss function, 
plasmon decay rate, plasmon oscillator strength, absorption spectral function and plasmon oscillation 
ratio to understand the physical characteristics of plasmons in ML-MoS$_2$.

\subsection{Optical absorption of circularly polarized light}
\begin{figure}[t]
\includegraphics[width=8.6cm]{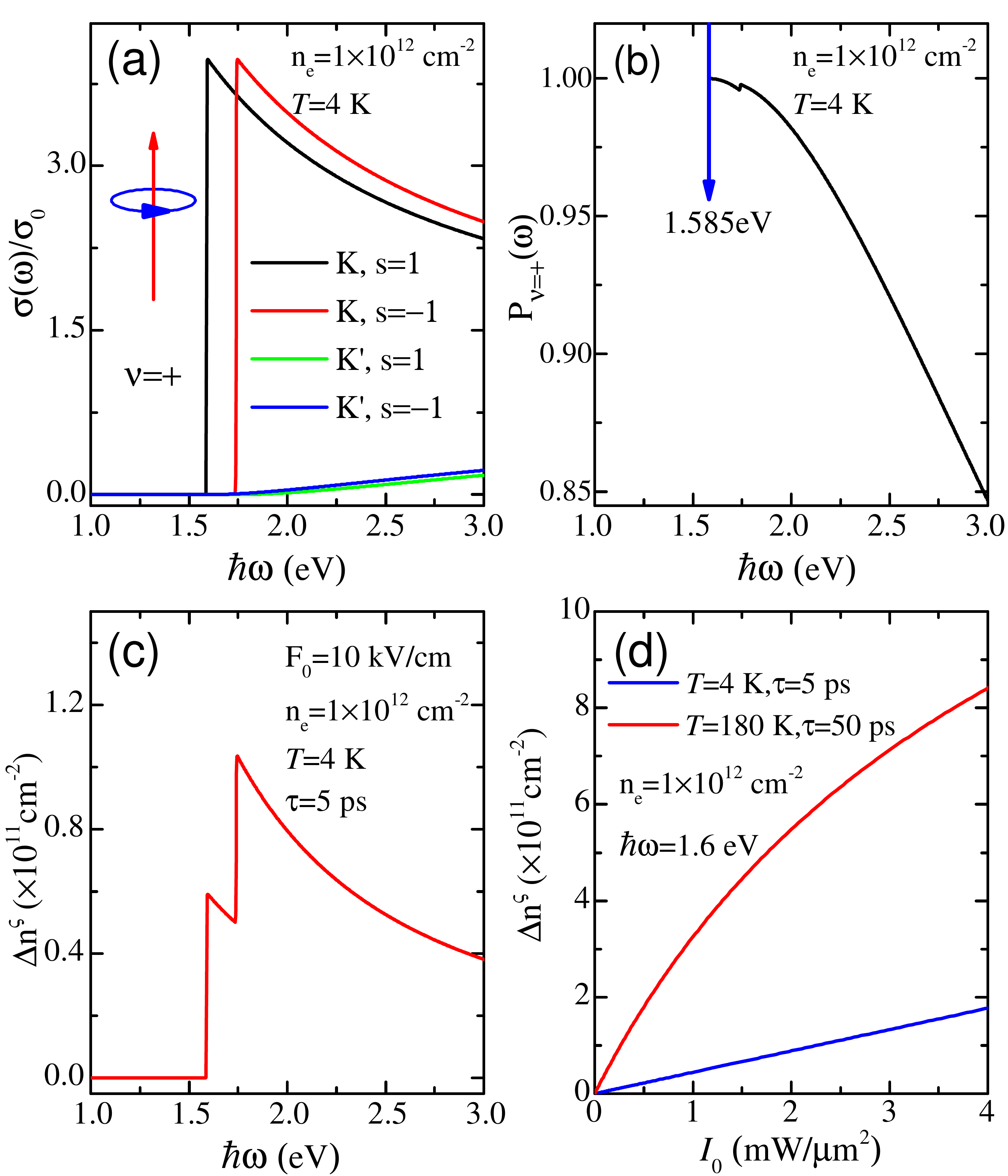}
\caption{(Color online) (a) Contributions to the inter-band optical conductivity
$\sigma^{\varsigma s}_\nu(\omega)$ for different valley and spin subsystems and (b) the
VA  of right-handed ($\nu=+1$) circularly polarized radiation by ML-MoS$_2$ as a function
of photon energy. (c) Photo-excited carrier density due to right-handed circularly polarized
light pumping at the $K$ valley as a function of photon energy. The sudden increase is due to
the contribution from the spin-down subsystem. (d) Photo-excited carrier density at $K$ valley
as a function of the radiation intensity $I_0\sim F^2_0$ of right-handed circularly polarized light
with fixed photon energy, where $F_0$ is the electric field strength of the incident light.}\label{fig2}
\end{figure}

In Fig. \ref{fig2}(a), we show the inter-band optical conductivity of an $n$-doped system
generated by species of electrons with different valley and spin in response to a right-handed
circularly polarized light beam ($\nu=+$). Correspondingly, the valley-dependent absorption (VA) 
is depicted in Fig. \ref{fig2}(b). The electron density is $n_e=1\times10^{12}$ cm$^{-2}$,
the system is assumed to be at low temperature, i.e. $T=4$ K and the lifetime of the photo-excited carriers
is $\tau=$ 5 ps \cite{Korn11}. Fig. \ref{fig2}(a) shows that the optical conductivity,
and hence the corresponding optical absorption, is the largest in the $K$ valley when the system
is illuminated with right-handed polarized light. When the photon energy is near the edge of the band gap, absorption within the spin-up subband at the $K$ valley plays a dominant role while the absorption
within the spin-down subband is suppressed. This is due to large spin splitting in the valence
band. The VA in Fig. \ref{fig2}(b) decreases with increasing photon energy as more transitions
become available. Notice the sudden increase in the VA curve which is due to the contribution
from the inter-band transitions within the spin-down subband at the $K$ valley. We notice that one
could obtain nearly 100$\%$ VA for a circularly polarized light with a photon frequency near the
band gap \cite{Kioseoglou12}.

In Fig. \ref{fig2}(c), we show the photo-excited carrier density under right-handed circularly
polarized light at the $K$ valley as a function of photon energy for a fixed strength of the electrical
field. The photo-excited carrier density in Fig. \ref{fig2}(c) is proportional to the sum of the
contributions to the spin and valley resolved optical conductivity presented in Fig. \ref{fig2}(a).
In Fig. \ref{fig2}(d), we show the photo-excited carrier density in the $K$ valley as a function
of the intensity of the incident radiation. In the presence of circularly polarized light
with photon energy $\hbar\omega=$ 1.6 eV, the VA is nearly 100$\%$ which means only electrons
in the $K$ valley are pumped. In the limit of weak light intensities, we observe that the 
photo-excited carrier density has a linear dependence on the radiation intensity, crossing over 
to a square root dependency for stronger radiation intensities when $T$=180 K. Notice that the 
photo-excited carrier life time varies with temperature \cite{Korn11} and that a larger carrier 
density can be obtained at higher temperature which can be seen from Eq. \eqref{photo-excited} 
and Fig. \ref{fig2}(d).

These results show that in order to efficiently excite charge carriers in a single valley, one has
to use a circularly polarized light beam with photon energy close to the bandgap. This will break
valley degeneracy in the ML-MoS$_2$ system and make it a multi-component system. The plasmonic
response of these systems is discussed in the next sections.

\subsection{Comparison between massive and hyperbolic description}

\begin{figure}[t]
\includegraphics[width=7.0cm]{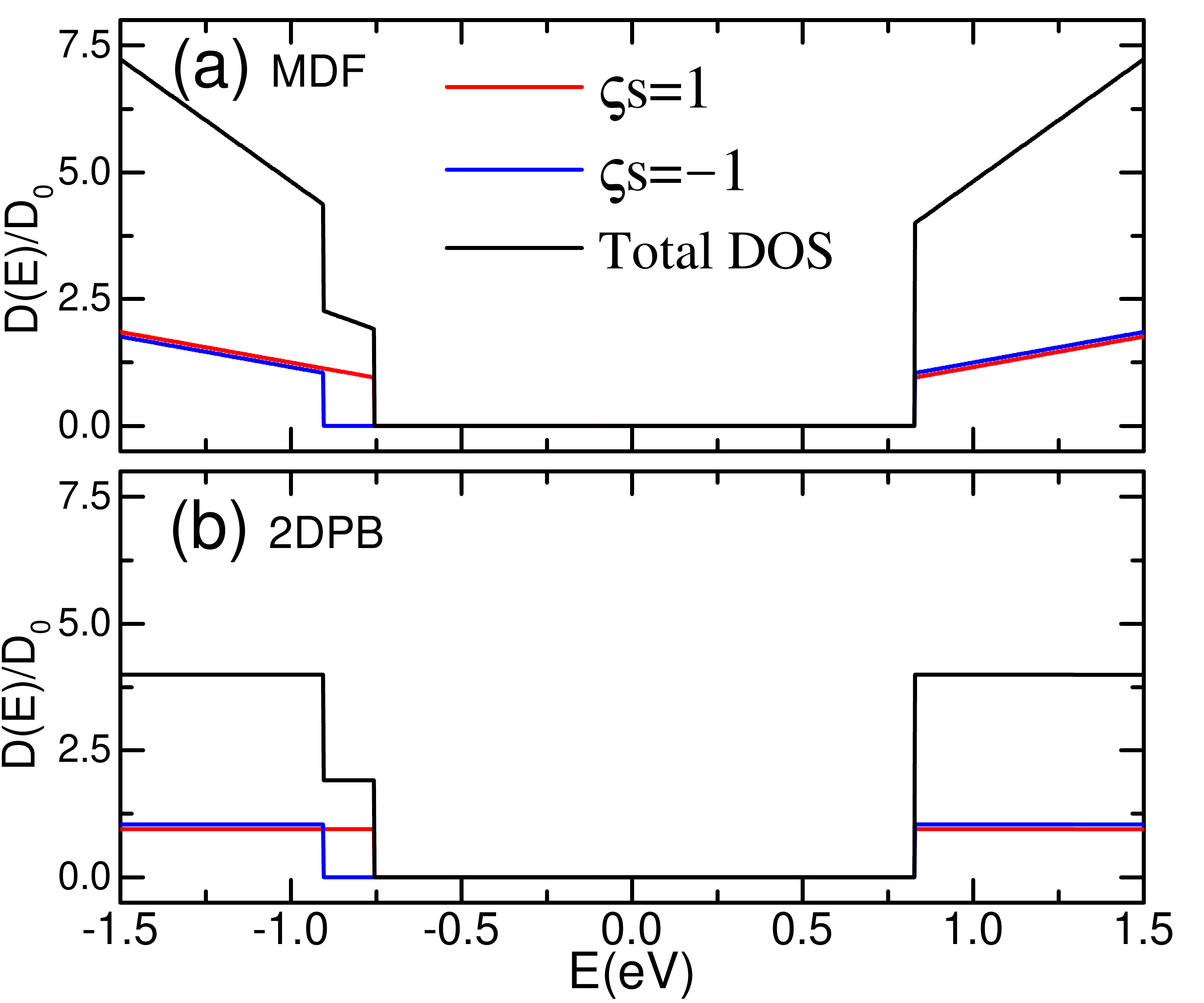}
\caption{(Color online) Density of states for charge carriers in ML-MoS$_2$ with (a)
MDF model and (b) 2DPB model, respectively. Here we used the notation $D_0=\Delta/(4\pi a^2t^2)$.}
\label{fig3}
\end{figure}

Usually, the band structure of ML-MoS$_2$ is described by a massive Dirac Fermion (MDF)
model \cite{Xiao12}. However, due to the large band gap in ML-MoS$_2$, the effective low-energy band structure of the MDF can also be approximately described by two-dimensional
electron/hole gases with two-dimensional parabolic bands (2DPB) if the carrier concentration 
in the bands is not too large. The relation between these two models was shown before in 
Eq. \eqref{2Dproxi}. In this section we show the equivalence of these two models in describing 
the optical response of ML-MoS$_2$.

In Fig. \ref{fig3}, we plot the density of states (DOS) of ML-MoS$_2$ for MDF and 2DPB models,
respectively. In the energy regime near the bottom/top of the conduction/valence band, the DOS of
the two models is very close, while as the energy increases, the two models deviate from each other.
This indicates that we can, indeed, use a 2DPB model to describe the band structure of ML-MoS$_2$ with 
low carrier density.\par

\begin{figure}[t]
\includegraphics[width=8.6cm]{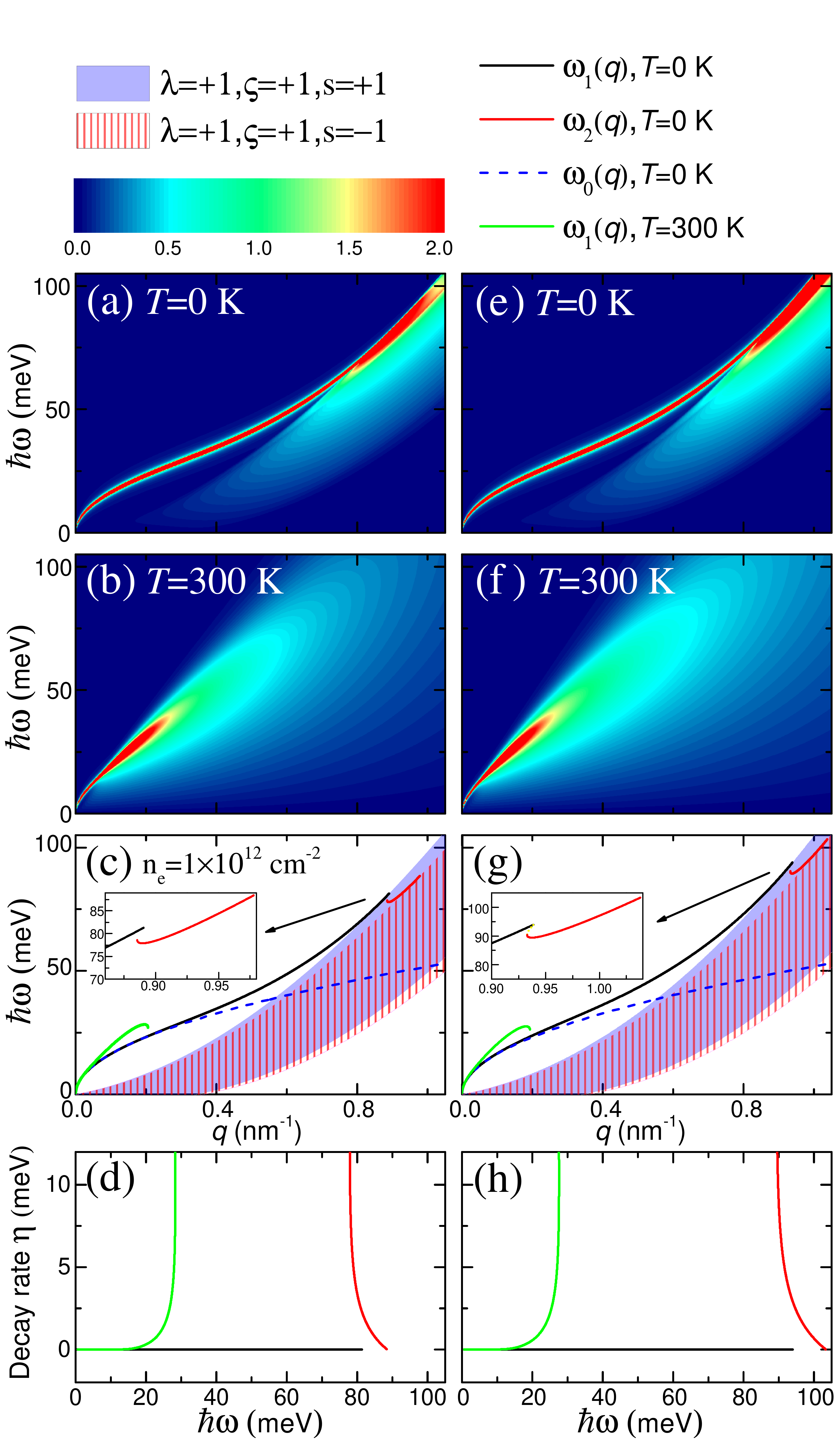}
\caption{(Color online) The energy loss function for the MDF model at (a) $T=0$ K and
(b) $T=300$ K. The plasmon dispersion and plasmon decay rate for the MDF model are shown
in (c) and (d), respectively. (e)-(h) are the the corresponding results of (b)-(e)
for the 2DPB model. The insets in (c) and (g) are a zoom of the large $q$ area as
indicated by the arrows. The blue dashed curve corresponds to the zero temperature
charge plasmon mode as obtained within the low-$q$ approximation. }\label{fig4}
\end{figure}

In Fig. \ref{fig4}, we compare the plasmonic properties calculated within the two models for an
$n$-type ML-MoS$_2$ whose band structure and occupation is shown in Fig. \ref{fig1}(a). We show the 
electron energy-loss function in the ($q$-$\omega$) plane at zero and room temperature, the plasmon 
dispersion and the decay rate of these modes. The electron density is fixed at $n_e=1\times10^{12}$ 
cm$^{-2}$ and equally distributed over the two valleys. In Figs. \ref{fig4}(a) and (e), we compare 
the zero-temperature energy-loss functions. The plasmon appears as a curve of strong absorption in 
the long-wavelength limit. For large $q$, the plasmon branch merges with the continuum of intra-band 
single-carrier excitations, which shows up as an increased absorption. Notice that the two models 
give qualitatively the same results, and also quantitatively they agree very closely.

Panels (b) and (f) of Fig. \ref{fig4} show the energy-loss function at room temperature. The
result shows that a finite temperature damps the plasmon, inhibiting collective excitations at
larger energies and wave vectors. Notice that also here the two models give qualitatively and
quantitatively very similar results.

In panels (c) and (g) of Fig. \ref{fig4} we show the roots of the real part of the dielectric
function $\epsilon_{\rm PRA}(\mathbf{q},\omega)$ as solid curves. The black curve is the 2D
charge plasmon, responsible for the thin line of enhanced absorption in the previous panels.
As shown by the dashed blue curve, in the long-wavelength limit this mode coincides with the
$\sqrt{q}$-plasmon mode from Eq. (\ref{square_root_plasmon}). The green solid curve in these
panels shows the plasmon mode for a finite-temperature system. The results show that at room 
temperature this mode is limited to the long-wavelength regime. Remarkably, in both panels, shown
by a solid red curve, there is also a new mode appearing inside the intra-band continuum of
the spin-up/spin-down particles at $K$/$K^{\prime}$ valley. This new mode is responsible for the 
enhanced values of the electron-energy loss function in panels (a) and (e) and is not appearing only
due to a breaking of the spin-degeneracy.

In panels (d) and (h) we show the decay rate as a function of the photon energy
for each mode. The black solid curve at the bottom of the panel refers to the normal charge
plasmon mode. The decay rate for this mode is zero for every energy because it lies outside
the particle-hole continuum. The newly found mode, shown by the red curve, however, is partly
Landau damped and acquires a finite decay rate. Notice that the decay rate decreases for
larger energy and is of the same order of magnitude as that of the normal plasmon mode at
room temperature, shown by the green curve in Figs. \ref{fig4}(d) and (h).

\begin{figure}[t]
\includegraphics[width=8.5cm]{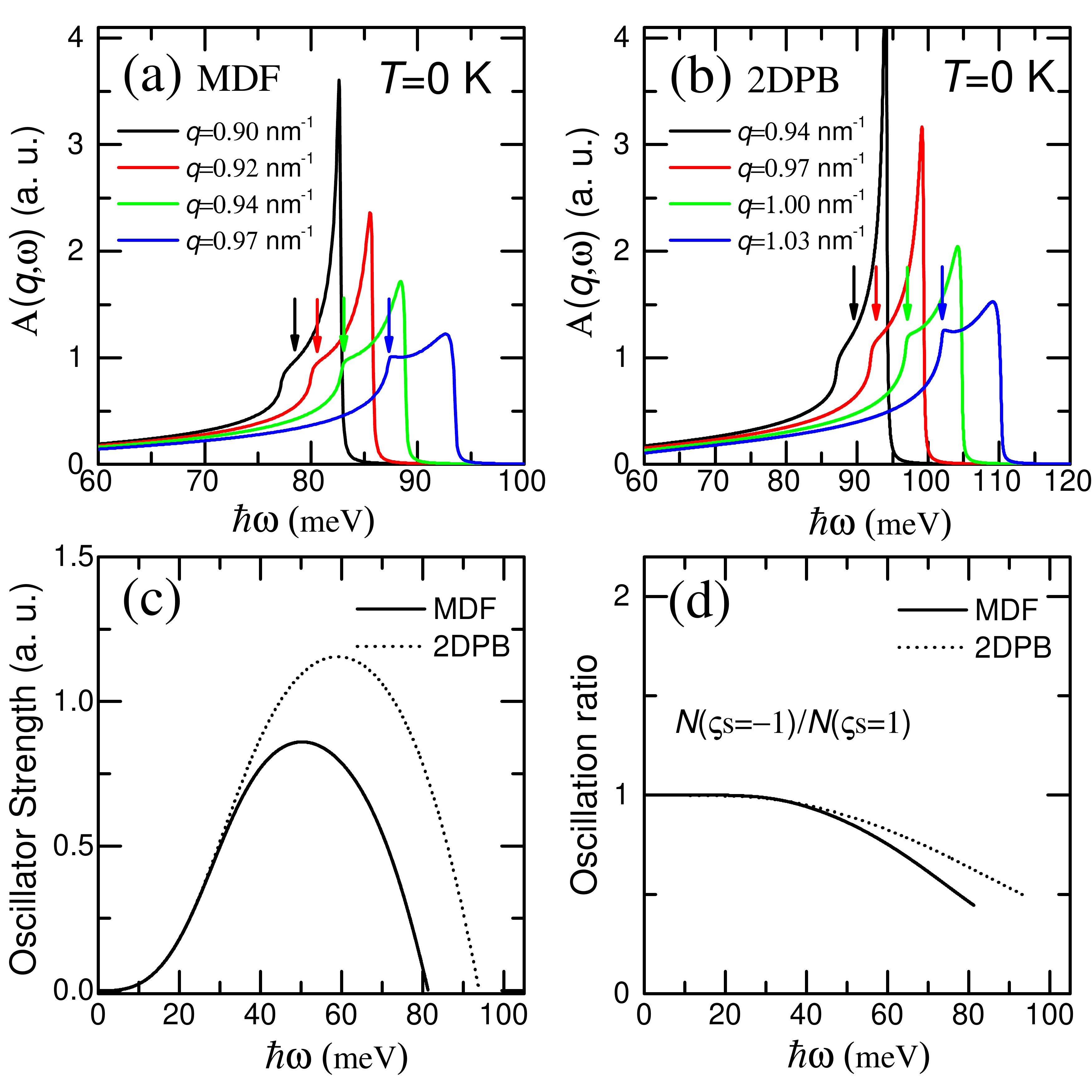}
\caption{(Color online) The absorption spectral function of an $n$-type ML-MoS$_2$ for
the situation corresponding to Fig. \ref{fig4} as a function of energy for different fixed wavevectors
for (a) the MDF model and (b) the 2DPB model at $T=0$ K. (c) The oscillator strength of the zero
temperature undamped plasmon for the two models. (d) The oscillation ratio of the two different
components in $n$-type ML-MoS$_2$ for the two models. The arrows in (a)-(b) indicate the
corresponding plasmon frequency for the second plasmon branch $\omega_2(q)$.}\label{fig5}
\end{figure}

In order to characterize the new, partly damped plasmon mode further, in Fig. \ref{fig5},
we show the spectral function $A(q,\omega)$ for different values of the wave vector $q$. We
see that the new mode appears as a shoulder to the spectral function that should be
distinguished from the peak in the spectral function at the edge of the particle-hole
continuum. Panels (a) and (b) of Fig. \ref{fig5} show that the MDF and 2DPB models have
a qualitative correspondence, but quantitatively they differ slightly. This difference
also surfaces when calculating the oscillator strength and the ratio of the amplitudes in both
valleys in panels (c) and (d), respectively. It shows that the 2DPB model overestimates the
oscillator strength and the plasmon frequency for a given wave vector. However, the qualitative
behaviour is similar. We can, therefore, safely use the 2DPB model to investigate
the plasmonic properties in ML-MoS$_2$ bearing in mind that the results might be quantitatively
slightly departing.

\subsection{Spin-polarized two-component system at finite temperature}

\begin{figure}[t]
\includegraphics[width=8.6cm]{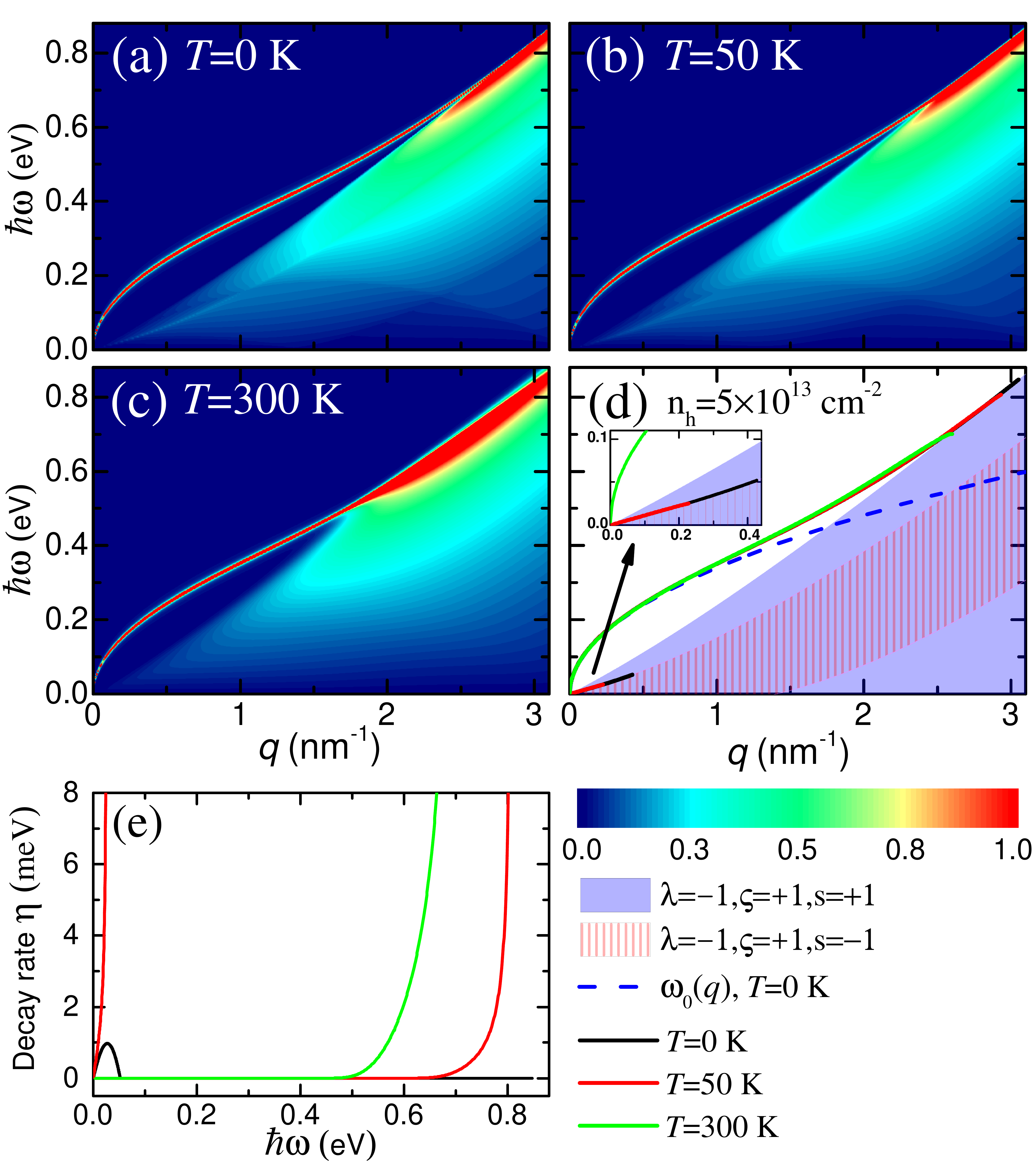}
\caption{(Color online) (a)-(c) The energy loss function of strongly doped $p$-type ML-MoS$_2$
in the $(q,\omega)$ plane for different temperatures. (d) The plasmon dispersion as a
function of wavevector for the different temperatures. (e) The plasmon decay rate as a
function of plasmon energy for the plasmon modes in (c). The inset in (d) is the zoom  of the small $q$ and
$\omega$ regime as indicated by the arrow.}\label{fig6}
\end{figure}

We now turn to the discussion of a two-component system where the two components are characterized
by a different spin. We can obtain this system in MoS$_2$ by tuning the Fermi level such
that the two spin resolved valence bands are occupied by holes as depicted in Fig. \ref{fig1}(c).
Notice that because in the opposite valley the upper and lower valence bands are reversed, there
is no macroscopic spin imbalance. Nonetheless, the spin-imbalance in a single valley has an effect 
on the optical and plasmonic properties of the system as will be shown below.

In Fig. \ref{fig6} we show the energy loss function, plasmon dispersion and the plasmon decay
rate for $p$-type ML-MoS$_2$ using the MDF model at a fixed hole density $n_h=5\times10^{13}$
cm$^{-2}$. We show results for three different temperatures. The energy loss functions and
corresponding plasmon dispersions at each temperature shows that there is a charge plasmon mode
in the small $q$ regime. Notice, however, that in contrast to the $n$-type system, the plasmon
dispersion is not affected a lot by temperature as is shown in Fig. \ref{fig6}(d). In panel (e),
we show the decay rate of the plasmon modes at each temperature and find that also at finite
temperature the lifetime is still appreciable, in strong contrast to $n$-type ML-MoS$_2$ discussed 
in the previous section.

The plasmon branches in Fig. \ref{fig6}(d) also reveal an additional peculiarity in the
long-wavelength limit. Indeed, there exists a weakly damped linear acoustic plasmon mode in
between the upper boundaries of the intra-band PHES of the two valence subbands. In this region the 
PHES has a local minimum and it is expected that the mode is, therefore, relatively stable. The distinct 
acoustic mode is similar to the one previously discussed in general spin-polarized two-dimensional 
electron gases \cite{Agarwal14,Kreil15}, but is now present in the absence of a macroscopic spin-imbalance and investigated at non-zero temperature.

\begin{figure}[t]
\includegraphics[width=8.6cm]{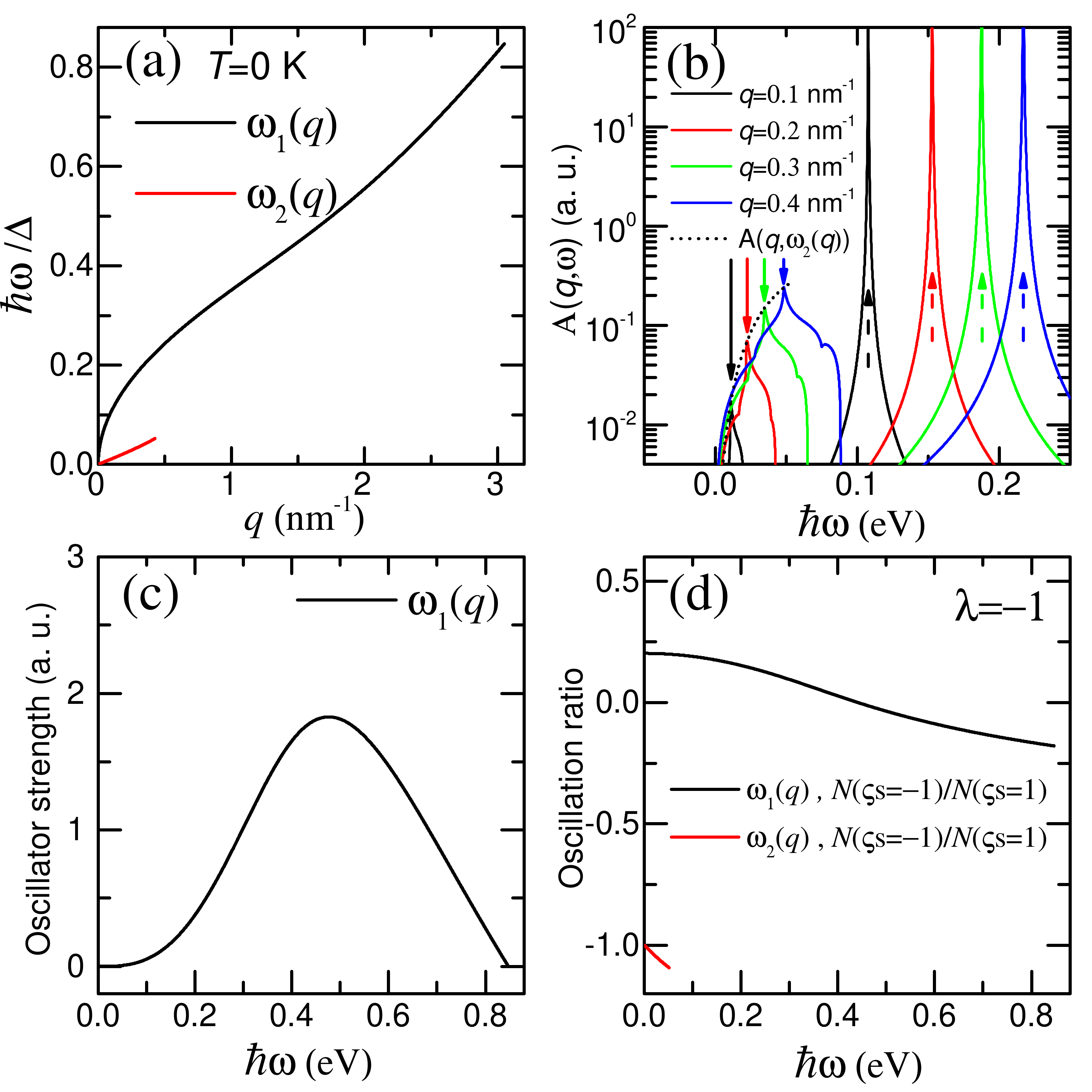}
\caption{(Color online) (a) The plasmon dispersion as a function of wave vector for
$p$-type ML-MoS$_2$ at $T=0$ K as shown in Fig. \ref{fig6}(d). (b) The absorption
spectral function as a function of energy $\omega$ for fixed wavevectors and the
absorption spectral function $A(q,\omega_2(q))$ (dashed line) for the linear acoustic
plasmon $\omega_2(q)$. The solid and dashed arrows correspond to the weakly damped
linear acoustic plasmon and undamped charge plasmon energies for fixed wavevectors.
(c) Plasmon oscillator strength of the undamped plasmon mode $\omega_1(q)$. (d) Ratio
of the plasmon oscillation amplitudes for the two spin components for the the two
plasmon modes as indicated.}\label{fig7}
\end{figure}

In Fig. \ref{fig7}(a), we show the plasmon dispersion of a $p$-type ML-MoS$_2$ for
the MDF model with a fixed hole density $n_h=5\times10^{13}$ cm$^{-2}$ as a function
of wave vector for the zero temperature case as in Fig. \ref{fig6}(d). From this plot,
one can clearly identify the normal $\omega_{1}(q)$ plasmon mode, and the acoustic
mode $\omega_{2}(q)$. In Fig. \ref{fig7}(b), we plot the spectral function for fixed
plasmon wave vectors. Notice that the linear plasmon mode appears as a peak on top of
the background particle-hole weight and, therefore, is expected to be less clear than
the distinct $\omega_{1}(q)$ plasmon. In panel (c) of Fig. \ref{fig7}, we show the oscillator
strength of the $\omega_{1}(q)$ plasmon mode, which shows a similar behaviour as observed
before.

To identify the character of the acoustic mode, in Fig. \ref{fig7}(d), we
plot the ratio of the amplitude of the oscillation for both spin components. We see that
for the new mode, displayed in red, this ratio is negative and approaches $-1$ in the
long-wavelength limit. This means that the density of the two spin components oscillates
in anti-phase. Therefore, this mode was labeled before as a spin-plasmon because this means that the carrier density does not oscillate, but it is rather the spin of the carriers that forms an oscillating pattern \cite{Agarwal14}.
As a consequence of the carrier density oscillation being in anti-phase for each spin
component, the mode is nearly charge neutral. Inspecting the $\omega_1(q)$ mode in panel
(d) shows that the two spin-components oscillate in-phase for small frequency, but then
cross-over to an oscillation in anti-phase. At this crossing point, the amplitude of the
oscillation in the spin-down hole liquid is zero, and hence, the plasmon is spin-polarized.

\subsection{Valley polarized three-component system}

\begin{figure}[t]
\includegraphics[width=8.6cm]{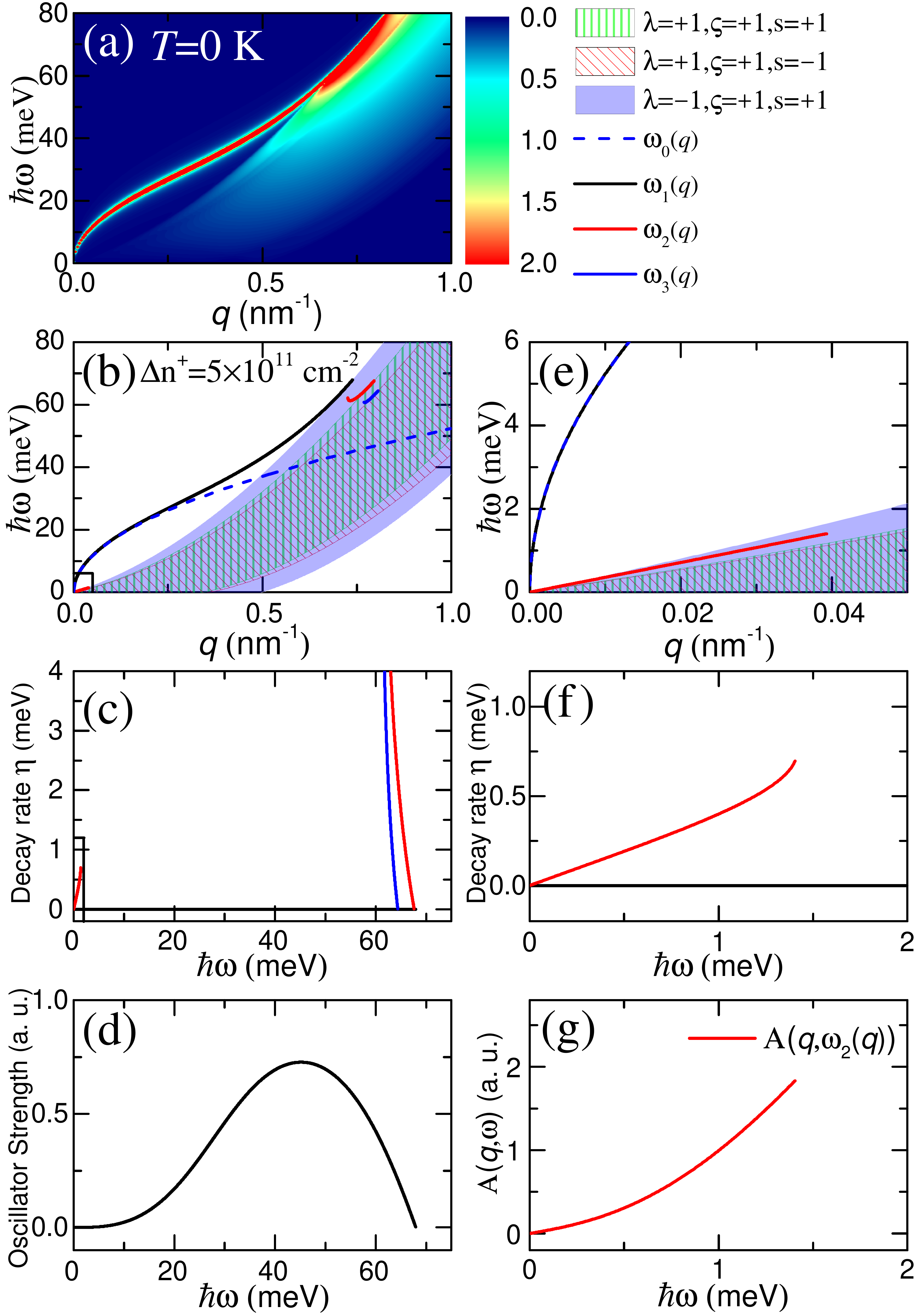}
\caption{(Color online) (a) The energy loss function and (b) plasmons
dispersion as a function of wave vector for an undoped ML-MoS$_2$ with a
photo-excited carrier density $\Delta n^+=5\times10^{11}$ cm$^{-2}$ by a right-handed
circularly polarized optical pumping. (c) The corresponding plasmon decay rate as a
function of plasmon energy. (e)-(f) are the zoom in of the black block areas in (c)-(d).
(d) The plasmon oscillator strength of the undamped plasmon $\omega_1(q)$ as a function
of plasmon energy. (g) The absorption spectral function of the weakly damped linear
acoustic plasmon model $\omega_2(q)$.}\label{fig8}
\end{figure}

In Fig. \ref{fig8}, we show the plasmonic behaviour of an undoped ML-MoS$_2$ sheet with
a photo-excited carrier density $\Delta n^+=5\times10^{11}$ cm$^{-2}$ by right-handed
circularly polarized light at zero temperature. In Fig. \ref{fig1}(d), the band structure
and carrier occupation are shown. In this system, free carriers exist only in the $K$ valley
thanks to the valley-dependent photo-excitation. In the $K^{\prime}$  valley, no
free carriers exist and, therefore, no plasmon propagation is possible.

This system can be regarded as a three-component system. Indeed, only one spin valence
band is depleted by the photo-excitation, generating a single hole liquid, but because
of very fast spin relaxation in the conduction band, both spin bands are occupied. This
generates two independent free carrier liquids. For this system, the energy loss function
in the ($q$-$\omega$) plane is shown in Fig. \ref{fig8}(a) and the plasmon dispersion as a
function of $q$ is shown in Fig. \ref{fig8}(b). The plasmon branch shows up clearly in
Fig. \ref{fig8}(a) and corresponds to an undamped charge plasmon mode. In Fig. \ref{fig8}(b),
we can see that there exist now in total three plasmon modes in three regions divided by
the upper boundaries of the intra-band PHES for the three components. These additional modes
are responsible for the increased energy loss in that region in the $q-\omega$ plane. The corresponding plasmon
decay rates are shown in Fig. \ref{fig8}(c) and show that the new plasmon modes have indeed
a finite lifetime. The undamped charge plasmon mode $\omega_1(q)$, however has zero decay
rate.

In the small $q$ limit, there is a weakly damped linear acoustic plasmon mode which we label by $\omega_2(q)$.
To show this mode more clearly, in Figs. \ref{fig8}(e)-(f) we show a zoom of the black box in Figs. \ref{fig8}(b)-(c). Fig. \ref{fig8}(d) shows that the plasmon oscillator strength for $\omega_1(q)$ increases first 
and then decreases to zero. For the weakly damped linear acoustic plasmon mode $\omega_2(q)$ in 
the small $q$ regime, the plasmon decay rate and the strength of its absorption spectral function 
increase with increasing plasmon energy.

In Sec. \ref{sec:theoretical approach C}, we obtained the polarization function for
each valley and spin subsystem of the MDF model which contains the contribution from both the
intra-band transitions within the conduction and valence bands and the inter-band transitions
between the conduction and valence bands. Thus, we cannot directly calculate the plasmon
oscillation ratio of the different subband components through Eq. \eqref{Osratio} with the
polarization function of the MDF model for a photo-excited system. Instead, we resort to
the description of the system with a 2DPB model.

\begin{figure}[t]
\includegraphics[width=7cm]{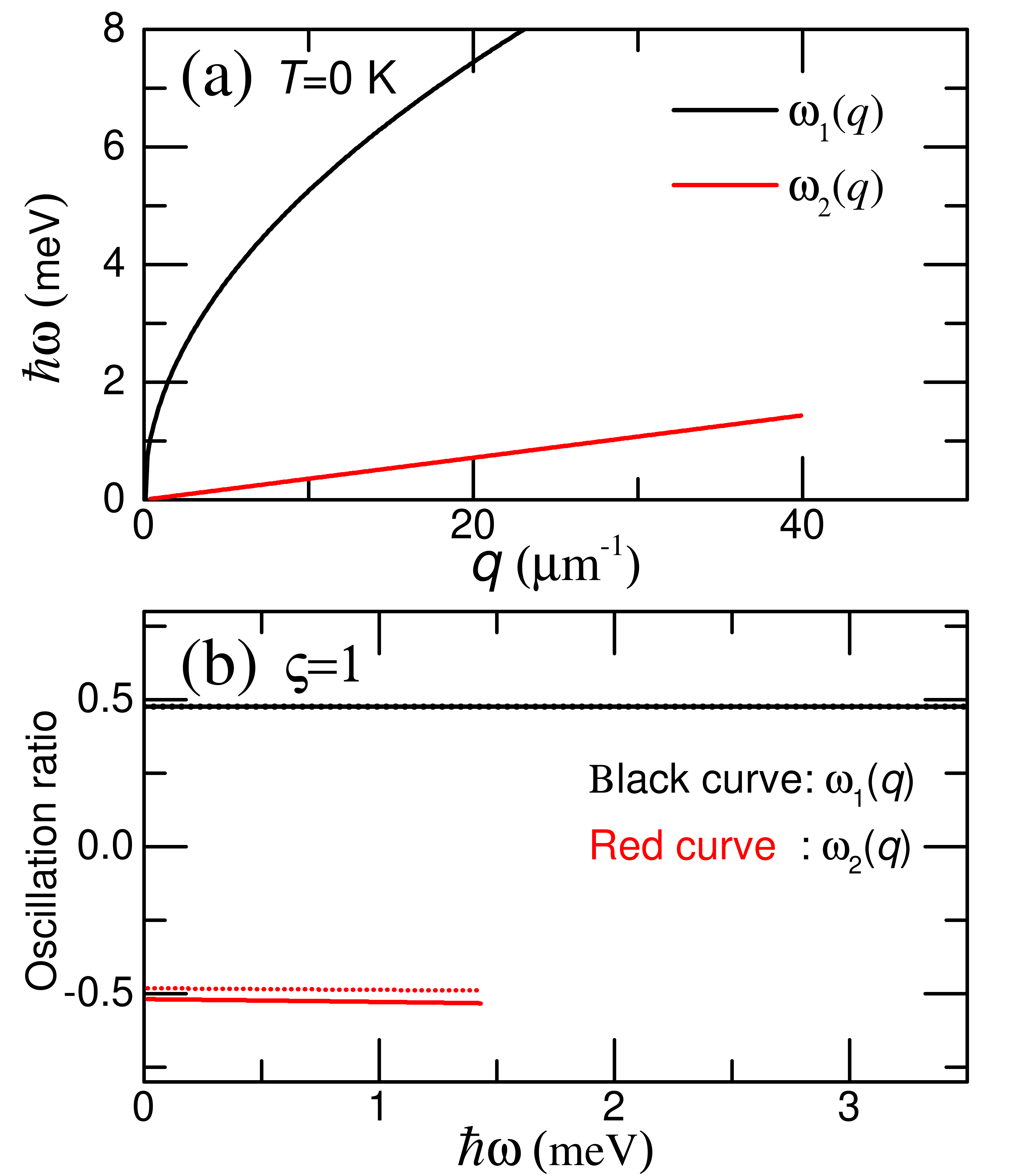}
\caption{(Color online) (a) The plasmon dispersion within the 2DPB model of a photo-excited undoped 
ML-MoS$_2$. The carrier density $\Delta n^+=5\times10^{11}$ cm$^{-2}$ is induced by right
handed circularly polarized light pumping. (b) The oscillation ratio between the spin-up 
conduction subband at the $K$ valley and the spin-up valence subband at the $K$ valley 
(solid lines) and between the spin-down conduction subband at the $K$ valley and the spin-up 
valence subband at $K$ valley (dashed lines). The black and red curves correspond to the 
results of plasmon modes $\omega_1(q)$ and $\omega_2(q)$, respectively.}\label{fig9}
\end{figure}

In Fig. \ref{fig9}, we show the plasmon dispersion and plasmon oscillation ratio for the
corresponding situation shown in Fig. \ref{fig8} but now with the 2DPB model. We find that
the plasmon dispersion of 2DPB model in Fig. \ref{fig9}(a) coincides with the results in
Fig. \ref{fig8}(e). Within the 2DPB model, we can calculate the ratio's of the amplitude of 
oscillation in three different components considered in this system. In Fig. \ref{fig9}(b), 
we show these oscillation ratios as a function of the photon frequency. In these calculations, 
the amplitude of the oscillation is calculated with respect to the amplitude of the valence 
band component. For the undamped plasmon mode $\omega_1(q)$, the spin-up and spin-down 
components in the conduction band and the spin-up component in valence band at $K$ valley 
oscillate in phase. Both electron components have an equally large oscillation. For the linear 
plasmon mode $\omega_2(q)$, however, the oscillation ratio between the conduction band and 
valence band is negative so the oscillation is in anti-phase. In the low-$q$ limit, the summation 
of the oscillation ratio for the acoustic plasmon mode $\omega_2(q)$ with the spin-up and down 
conduction subbands is equal to $-1$. This means that the strength of the oscillation in both 
electron components is also nearly equal, but in anti-phase with the hole liquid. This type of 
anti-phase behavior was noted before in spin-polarized 2D electron systems\cite{Agarwal14}. Now, 
however, the weakly damped linear plasmon is not charge neutral since the anti-phase oscillations 
occur in oppositely charged liquids of electrons and holes.

\subsection{Four-component system}

\begin{figure}[t]
\includegraphics[width=8.6cm]{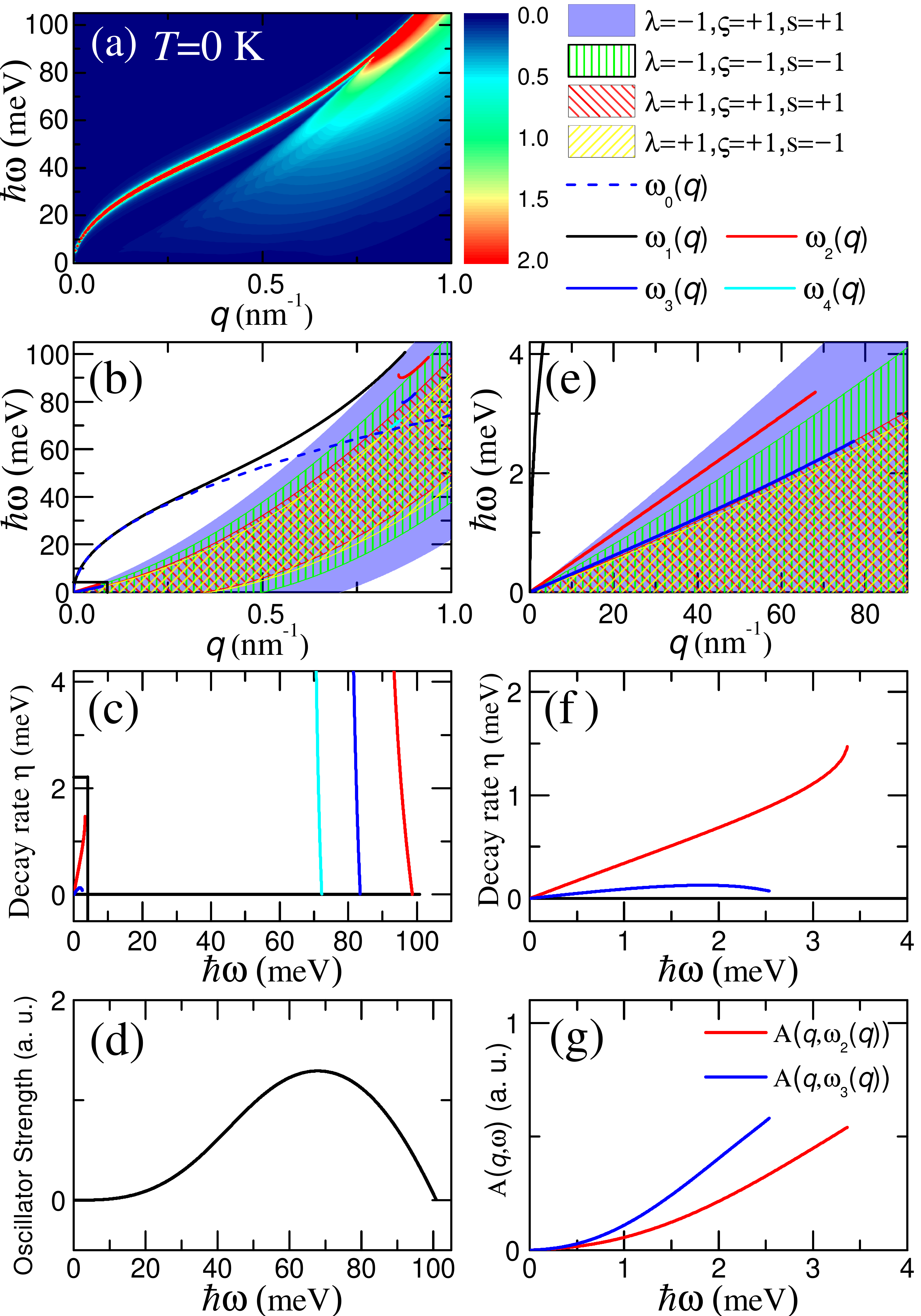}
\caption{(Color online) (a) The energy loss function and (b) the
plasmon dispersion as a function of wave vector for a photo-excited $p$-type
doped ML-MoS$_2$ with initial hole density $n_h=1\times10^{12}$ cm$^{-2}$ and a
photo-excited carrier density $\Delta n^+=5\times10^{11}$ cm$^{-2}$ induced by pumping with 
right-handed circularly polarized light. (c) The corresponding plasmon decay
rate as a function of plasmon energy. (e) and (f) are the zoom of the black boxes
in (b) and (c). (d) The plasmon oscillator strength of the undamped plasmon
$\omega_1(q)$ as a function of plasmon energy. (g) The absorption spectral function
of the two weakly damped linear acoustic plasmon modes.}\label{fig10}
\end{figure}

We end the analysis with the discussion of a four-component system. As shown in
Fig. \ref{fig1}(e), such a system can be created when a $p$-type ML-MoS$_2$ is pumped with
circularly polarized light. This will excite electrons in one valley into the conduction
band. As the spin relaxation time is very small \cite{Mak14,Song13}, the quasi equilibrium
formed in the system consists of an electron pocket in one valley, but with electrons of
two spin types with the same Fermi level, and two pockets of holes distributed over the
two valleys with different Fermi level.

We assume that the $p$-type doped ML-MoS$_2$ has an initial hole density $n_h=1\times10^{12}$ cm$^{-2}$.
The photo-excited carrier density is $\Delta n^+=5\times10^{11}$ cm$^{-2}$, which is induced by
right-handed circularly polarized optical pumping. The energy loss function in Fig. \ref{fig10}(a)
shows the typical plasmon pole corresponding to the $\omega_1(q)$ mode as in the previous systems.
In Figs. \ref{fig10}(b) and (e) we, however, find that there exist three new plasmon branches for
large photon energy, and two new acoustic modes in the long-wavelength limit. As before, these new
modes appear in regions of the intra-band continuum where only some of the components are subject 
to Landau damping. The quantification of the importance of Landau damping is shown by the decay rate 
calculated in panels (c) and (f) where it is shown that, remarkably, the acoustic $\omega_{3}(q)$
mode is more stable than the acoustic $\omega_{2}(q)$ mode even though it can be found deeper in 
the intra-band electron-hole continuum. Furthermore, as shown in panel (g),
the $\omega_{3}(q)$ mode is more pronounced in the spectral function than the $\omega_{2}(q)$, so
one expects the former to be more easily excited. Finally, the oscillator strength calculated in
panel (d) of Fig. \ref{fig10} shows no qualitative differences due to the presence of the new modes
in the intra-band continuum. This means that these modes are the consequence of a redistribution of 
spectral weight inside the continuum rather than getting it from the regular plasmon.

\begin{figure}[t]
\includegraphics[width=7cm]{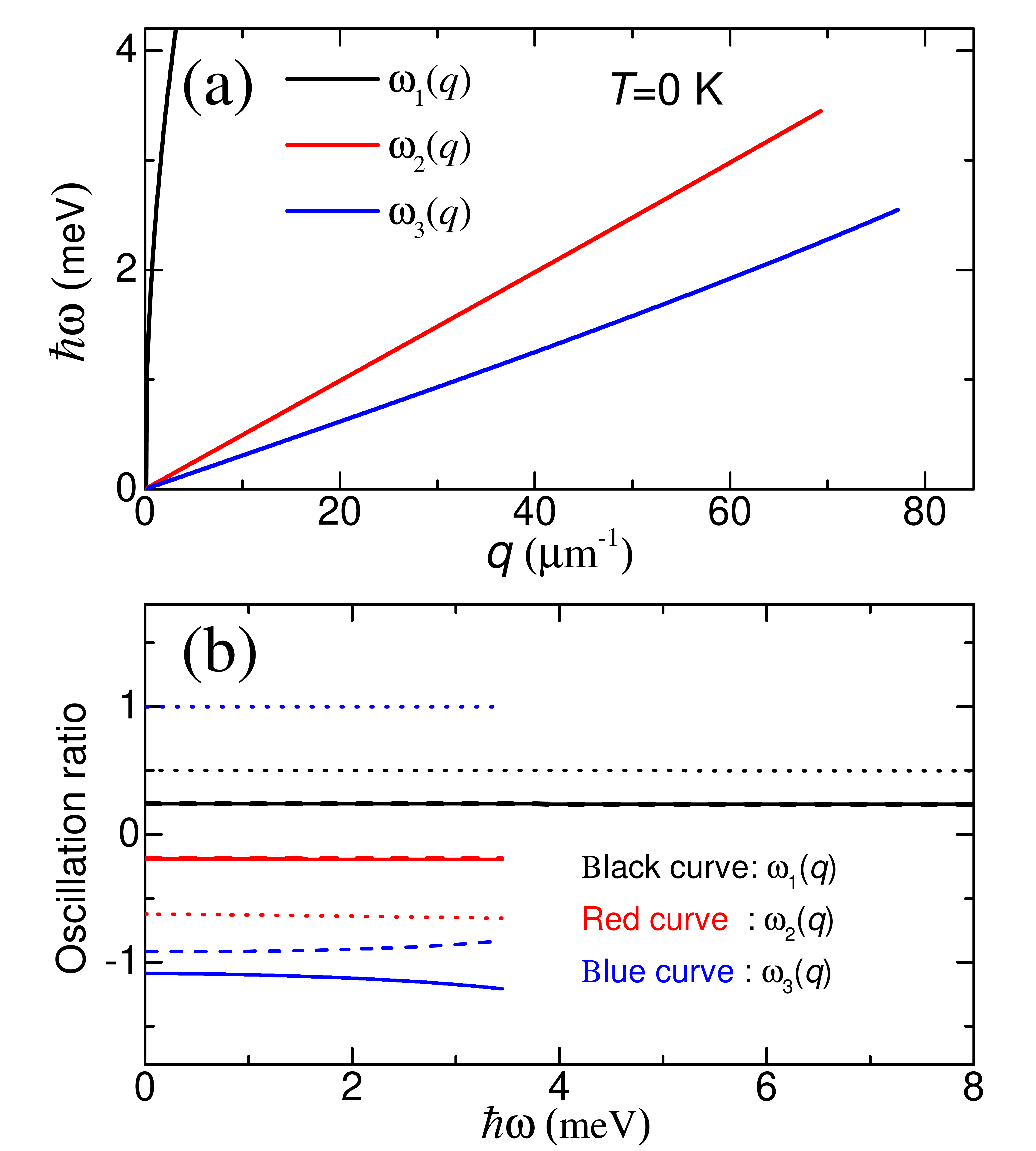}
\caption{(Color online) (a) The zero-temperature plasmon dispersion within the 2DPB model of an 
optically pumped $p$-type doped ML-MoS$_2$ with a initial hole density $n_h=1\times10^{12}$ cm$^{-2}$ and
a photo-excited carrier density $\Delta n^+=5\times10^{11}$ cm$^{-2}$ realized trough pumping with right-handed polarized light. (b) The ratio of the oscillation amplitudes in the spin-up conduction subband and the 
spin-up valence subband at $K$ valley (solid lines), the ratio between the spin-down conduction
subband and the spin-up valence subband at $K$ valley (dashed lines) and the ratio between the 
spin-down valence subband at $K^{\prime}$  valley and the spin-up valence subband at $K$ valley 
(dotted lines). The black, red and blue curves correspond to the results of the plasmon modes 
$\omega_1(q)$,  $\omega_2(q)$ and $\omega_3(q)$, respectively.}\label{fig11}
\end{figure}

In Fig. \ref{fig11}, we plot the plasmon dispersion and the oscillation ratio between the
different components for the long-wavelength acoustic modes discussed in the previous paragraph.
For these results, we investigated the system within the 2DPB model. We find the same
long-wavelength modes as before as we show in panel (a).

In panel (b), we calculate the ratio between the amplitudes of the different components in
the system for the three different plasmon modes. The solid curves denote the ratio between
the amplitude of the oscillations in the spin-up electron pocket and the spin-up hole pocket
in the $K$ valley. The results show that the $\omega_{1}(q)$ mode consists of in-phase oscillations,
while for the two other modes the oscillations are in anti-phase. The dashed curves in the same
panel denote the ratio of the oscillation amplitude between the spin-down electron pocket
and the hole pocket in the $K$-valley. For this, we find similar results as before, rendering
the $\omega_{1}(q)$ mode in-phase, while the $\omega_{2}(q)$ and $\omega_{3}(q)$
modes are in anti-phase. Notice, however, that since we are now considering oscillations in the
density of oppositely charged particles, it is the in-phased plasmon that compensate each
others charged oscillation. Finally, the dotted curves show the ratio between the hole pockets
in both valleys. Notice that, peculiarly, the $\omega_{3}(q)$ mode is completely in-phase and
that the amplitude for both components is the same in this case. The $\omega_{1}(q)$ is also
in-phase, as before, and the $\omega_{2}(q)$ mode is in anti-phase.

\section{Conclusions}
\label{sec:conclusions}
In this study, we examined the plasmonic response of a multi-component system. We took as
a platform a ML-MoS$_2$ system subjected to circularly polarized light. We have shown that
this system is capable of supporting multiple plasmon modes and we have quantified the effect of
circular optical pumping within a Boltzmann framework.

We found that all platforms support a long-wavelength $\sqrt{q}$ plasmon branch at zero
Kelvin. As temperature increases, we have shown that for an electron-doped system, the
plasmon dispersion is strongly affected, while for a hole-doped system this mode is much
more stable.

The main influence of having multiple components in the system is the appearance of distinct
plasmonic modes that are partly damped. We found that for an $n$-component system,
$n-1$ plasmon modes appear. These distinct modes manifest themselves in the long-wavelength
limit as acoustic modes with a linear dispersion and are in a local minimum of the intra-band
continuum. To evaluate their stability, we have calculated the decay rate for each of
these modes and found that although they lie in the intra-band continuum, their lifetime is 
considerable. Furthermore, for larger wave vectors, we found that the multi-component system also supports
new modes in regions where only some of the components are subject to Landau damping of the
collective oscillation.

We evaluated the character of the modes by investigating the ratios of the amplitudes
of the collective density oscillations of the different components. We found that the
oscillation for the regular $\sqrt{q}$-mode is in-phase, while for the acoustic modes
anti-phase oscillation is possible. Finally, we evaluated the spectral function,
showing how the acoustic modes can be identified.

In the course of the investigation, we have shown that in the parameter range suitable for MoS$_2$ with optical pumping, the massive Dirac Fermion model and the 2D parabolic band model yield very similar results for the plasmonic response. This is because inter-band transitions do not to affect the plasmons very strongly.

Recently, experiments on the plasmonic response of MoS$_2$ have been performed using electron energy loss spectroscopy \cite{Wang2015} and angle-resolved reflectance spectroscopy \cite{Liu2016}. Also phonon-plasmon modes have been investigated using Fourier-transform infrared spectroscopy \cite{Patoka2016}. Applying these techniques to ML-MoS$_2$ with circularly polarized optical pumping one should be able to differentiate between the distinct plasmon modes and quantify their lifetime and oscillator strength. 

The characteristic energy scale for plasmon modes in ML-MoS$_2$ covers not only the infrared 
but also the THz bandwidth, especially for the low frequency weakly damped linear acoustic 
plasmon modes. These properties make ML-MoS$_2$ a promising platform for plasmonic applications in
the infrared and THz frequency regime. The theoretical investigations in this paper will
help to guide the experimental search for new plasmon modes in ML-MoS$_2$ systems.

\section*{ACKNOWLEDGMENTS}
Y.M.X. acknowledges financial support from the China Scholarship Council (CSC).
B.V.D. is supported by the Flemish Science Foundation (FWO-Vl) by a postdoctoral fellowship. 
This work was also
supported by the National Natural Science Foundation of China (Grant No. 11574319,
11304272), Ministry of Science and Technology of China (Grant No. 2011YQ130018),
Department of Science and Technology of Yunnan Province, Applied Basic Research
Foundation of Yunnan Province (2013FD003), and by the Chinese Academy of Sciences.

\section*{APPENDIX: ANALYTICAL EXPRESSION OF POLARIZATION FUNCTION}
We use $\omega$ and $k$ as dimensionless variables to represent
$\hbar\omega$ and $atk$ for notational simplification. For zero doping,
the polarization function for spin $s$ band at $\varsigma$ valley is
\cite{Pyatkovskiy09}
\begin{align}
\Pi^{\varsigma s}_{0,T=0}(\mathbf{q},\omega)
=&-\frac{q^2}{4\pi a^2t^2}\bigg\{\frac{\Delta_{\varsigma s}}{q^2-\omega^2}
+\frac{q^2-\omega^2-4\Delta^2_{\varsigma s}}{4|q^2-\omega^2|^{3/2}}\nonumber\\
&\times\bigg[\theta(q-\omega)\arccos{\frac{\omega^2-q^2+4\Delta_{\varsigma s}^2}
{q^2-\omega^2+4\Delta_{\varsigma s}^2}}\nonumber\\
&-\theta(\omega-q)\ln{\frac{(\sqrt{\omega^2-q^2}+2\Delta_{\varsigma
s})^2}{|\omega^2-q^2-4\Delta^2_{\varsigma s}|}}\bigg]\bigg\}\nonumber\\
&-i\frac{q^2\tilde{x}^2_{\varsigma s}\theta(\omega^2-q^2
-4\Delta^2_{\varsigma s})}{16a^2t^2|\omega^2-q^2|^{1/2}},
\end{align}
where $\tilde{x}^2_{\varsigma s}=2-x^2_{\varsigma s}$ and
$x_{\varsigma s}=\sqrt{|1+4\Delta^2_{\varsigma s}/(q^2-\omega^2)|}$.

For finite doping ($\lambda=+$ for $n$-type doping and $\lambda=-$ for $p$-type doping),
the polarization function for the spin $s$ band at $\varsigma$ valley is
\begin{align}\label{No0Polarization}
\Pi^{\varsigma s\lambda}_{1,T=0}(\mathbf{q},\omega)
=&-\frac{u^\lambda_{\varsigma s}}{2\pi a^2t^2}
+\frac{q^2}{16\pi a^2t^2|q^2-\omega^2|^{1/2}}\nonumber\\
&\times
\begin{cases}
iG_>(y_-)-iG_>(y_+),                                   &\mathrm{1A}\\
G_<(y_-)-iG_>(y_+),                                    &\mathrm{2A}\\
G_<(y_+)+G_<(y_-),                                     &\mathrm{3A}\\
G_<(y_-)-G_<(y_+),                                     &\mathrm{4A}\\
G_>(y_+)-G_>(y_-),                                     &\mathrm{1B}\\
G_>(y_+)+iG_<(y_-),                                    &\mathrm{2B}\\
G_>(y_+)-G_>(-y_-)-i\pi\tilde{x}^2_{\varsigma s},  &\mathrm{3B}\\
G_>(y_+)+G_>(-y_-)-i\pi\tilde{x}^2_{\varsigma s},  &\mathrm{4B}\\
G_0(y_+)-G_0(y_-).                                    &\mathrm{5B}
\end{cases}
\end{align}
where $y_{\pm}=(2u^\lambda_{\varsigma s}\pm\omega)/q$, and
\begin{align}
&G_<(x)=x(x_{\varsigma s}^2-x^2)^{1/2}+(x_{\varsigma s}^2-2)
\arccos{(x/x_{\varsigma s})},\nonumber\\
&G_>(x)=x(x^2-x_{\varsigma s}^2)^{1/2}+(x_{\varsigma s}^2-2)
\mathrm{arccosh}(x/x_{\varsigma s}),\nonumber\\
&G_0(x)=x(x^2+x^2_{\varsigma s})^{1/2}+(-x^2_{\varsigma s}-2)
\mathrm{arcsinh}(x/x_{\varsigma s}).\nonumber
\end{align}

\begin{figure}[t]
\includegraphics[width=8.0cm]{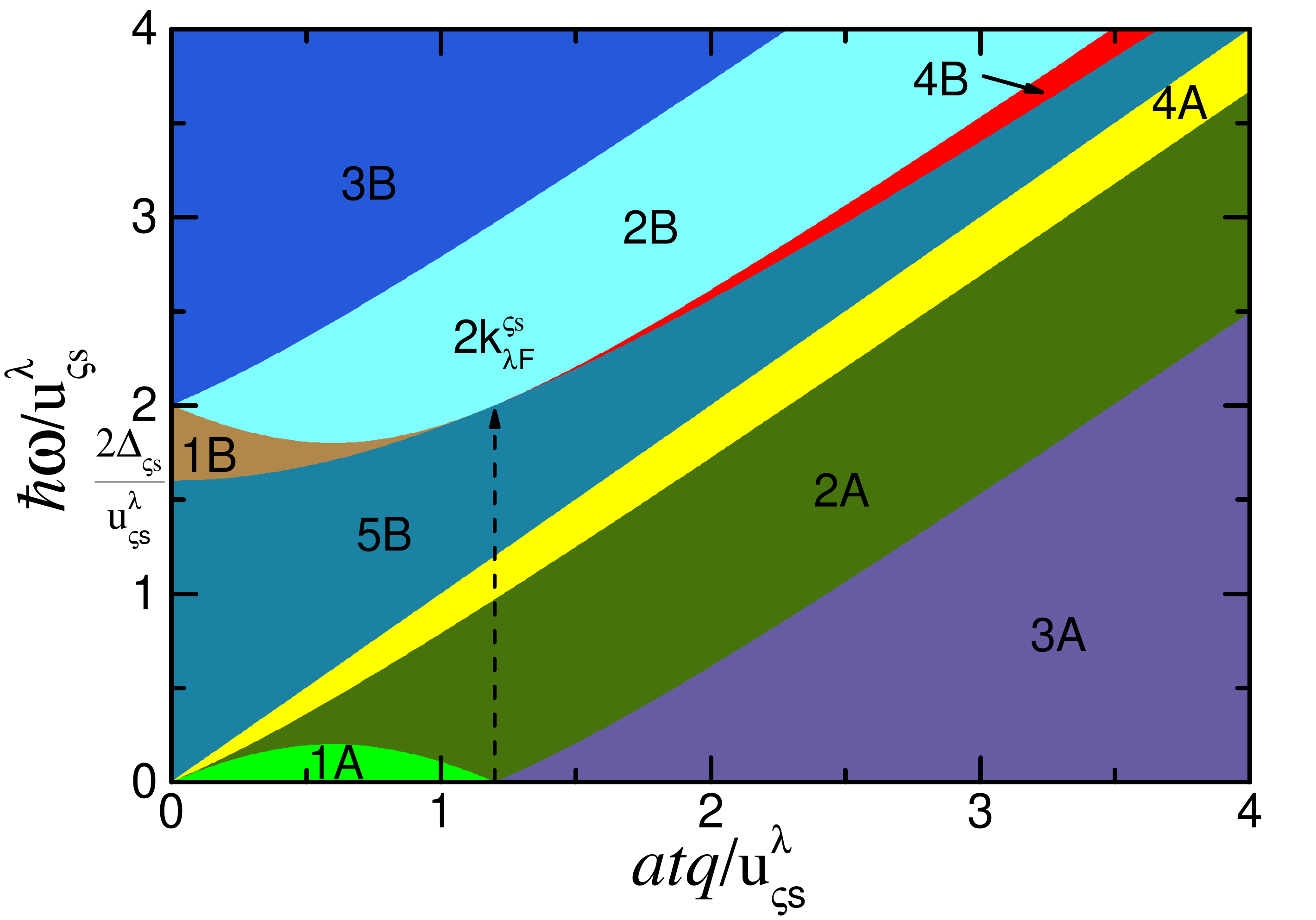}
\caption{(Color online) The different regions with different expressions for the
polarization function in Eq. \eqref{No0Polarization}. Here we have set
$\Delta_{\varsigma s}=0.8u^\lambda_{\varsigma s}$.}
\label{fig12}
\end{figure}

The regions defining the polarization function in
Eq. \eqref{No0Polarization} are defined as
\begin{align}
\mathrm{1A}:&\omega\leq u^\lambda_{\varsigma s}
-[(k^{\varsigma s}_{\lambda\mathrm{F}}-q)^2
+\Delta_{\varsigma s}^2]^{1/2},\nonumber\\
\mathrm{2A}:&\mp u^\lambda_{\varsigma s}\pm
[(k^{\varsigma s}_{\lambda\mathrm{F}}-q)^2
+\Delta_{\varsigma s}^2]^{1/2}<\omega\nonumber\\
&\leq-u^\lambda_{\varsigma s}+[(k^{\varsigma s}_{\lambda\mathrm{F}}+q)^2
+\Delta_{\varsigma s}^2]^{1/2},\nonumber\\
\mathrm{3A}:&\omega\leq-u^\lambda_{\varsigma s}
+[(k^{\varsigma s}_{\lambda\mathrm{F}}-q)^2
+\Delta_{\varsigma s}^2]^{1/2},\nonumber\\
\mathrm{4A}:&-u^\lambda_{\varsigma s}
+[(k^{\varsigma s}_{\lambda\mathrm{F}}+q)^2
+\Delta_{\varsigma s}^2]^{1/2}<\omega<q,\nonumber\\
\mathrm{1B}:&q\leq2k^{\varsigma s}_{\lambda\mathrm{F}},[q^2
+4\Delta_{\varsigma s}^2]^{1/2}<\omega\nonumber\\
&\leq u^\lambda_{\varsigma s}+[(k^{\varsigma s}_{\lambda\mathrm{F}}-q)^2
+\Delta_{\varsigma s}^2]^{1/2},\nonumber\\
\mathrm{2B}:&u^\lambda_{\varsigma s}
+[(k^{\varsigma s}_{\lambda\mathrm{F}}-q)^2
+\Delta_{\varsigma s}^2]^{1/2}<\omega\nonumber\\
&\leq u^\lambda_{\varsigma s}+[(k^{\varsigma s}_{\lambda\mathrm{F}}+q)^2
+\Delta_{\varsigma s}^2]^{1/2},\nonumber\\
\mathrm{3B}:&\omega>u^\lambda_{\varsigma s}
+[(k^{\varsigma s}_{\lambda\mathrm{F}}+q)^2
+\Delta_{\varsigma s}^2]^{1/2},\nonumber\\
\mathrm{4B}:&q>2k^{\varsigma s}_{\lambda\mathrm{F}},
[q^2+4\Delta_{\varsigma s}^2]^{1/2}<\omega\nonumber\\
&\leq u^\lambda_{\varsigma s}+[(k^{\varsigma s}_{\lambda\mathrm{F}}-q)^2
+\Delta_{\varsigma s}^2]^{1/2},\nonumber\\
\mathrm{5B}:&q<\omega\leq[q^2+4\Delta_{\varsigma s}^2]^{1/2}.\nonumber
\end{align}
where the Fermi vector $k^{\varsigma s}_{\lambda\mathrm{F}}=
\sqrt{|u^\lambda_{\varsigma s}|^2-\Delta_{\varsigma s}^2}$.

The polarization function for ML-MoS$_2$ calculated using the 2DPB model, 
Eq. \eqref{2Dproxi}, can be written as
\begin{equation}
\tilde{\Pi}(\mathbf{q},\omega)=\sum_{\varsigma s\lambda}
\tilde{\Pi}_\lambda^{\varsigma s}(\mathbf{q},\omega),
\end{equation}
where
\begin{equation}
\tilde{\Pi}_\lambda^{\varsigma s}(\mathbf{q},\omega)
=\sum_{\mathbf{k}}
\frac{f(\tilde{E}^{\varsigma s}_{\lambda\mathbf{k}})-
f(\tilde{E}^{\varsigma s}_{\lambda\mathbf{k+q}})}
{\omega+\tilde{E}^{\varsigma s}_{\lambda\mathbf{k}}
-\tilde{E}^{\varsigma s}_{\lambda\mathbf{k+q}}+i\delta}.
\end{equation}
At zero temperature, the real and imaginary part of the polarization
function $\tilde{\Pi}_{\lambda,T=0}^{\varsigma s}(\mathbf{q},\omega)$
can be written separately as \cite{Stern67,Giuliani05}
\begin{align}
\mathrm{Re} \tilde{\Pi}^{\varsigma s}_{\lambda,T=0}(\mathbf{q},\omega)
=&-\frac{\Delta_{\varsigma s}}{2\pi a^2t^2}\bigg[1+\sum_{\alpha=\pm}
\frac{\alpha \tilde{k}^{\varsigma s}_{\lambda \mathrm{F}}}{q}
\theta(|\nu^{\varsigma s}_\alpha|^2-1)\nonumber\\
&\times\mathrm{sgn}[\nu^{\varsigma s}_\alpha]
\sqrt{|\nu^{\varsigma s}_\alpha|^2-1}\bigg]
\theta(\tilde{u}^\lambda_{\varsigma s}
-\Delta_{\varsigma s}),\nonumber
\end{align}
and
\begin{align}
\mathrm{Im} \tilde{\Pi}^{\varsigma s}_{\lambda,T=0}(\mathbf{q},\omega)
=&-\frac{\Delta_{\varsigma s}}{2\pi a^2t^2}\sum_{\alpha=\pm}
\frac{\alpha \tilde{k}^{\varsigma s}_{\lambda \mathrm{F}}}{q}
\theta(1-|\nu^{\varsigma s}_\alpha|^2)\nonumber\\
&\times\sqrt{1-|\nu^{\varsigma s}_\alpha|^2}
\theta(\tilde{u}^\lambda_{\varsigma s}
-\Delta_{\varsigma s}),
\end{align}
where $\tilde{u}^\lambda_{\varsigma s}=|\tilde{E}^{\varsigma}_{\lambda\mathrm{F}}|
-\lambda\varsigma s\gamma/2$, the Fermi vector $\tilde{k}^{\varsigma s}_{\lambda\mathrm{F}}=
\sqrt{2\Delta_{\varsigma s}(\tilde{u}^\lambda_{\varsigma s}-\Delta_{\varsigma s})}$, and
\begin{equation}
\nu^{\varsigma s}_\alpha=\frac{\omega\Delta_{\varsigma s}}{q\tilde{k}^{\varsigma s}_{\lambda \mathrm{F}}}-\alpha\frac{q}{2\tilde{k}^{\varsigma s}_{\lambda \mathrm{F}}}.
\end{equation}

At zero temperature and in the low-$q$ approximation, the charge plasmon
dispersion of ML-MoS$_2$ within the 2DPB model is
\begin{equation}
\tilde{\omega}_0(q)=\big[e^2q\sum_{\lambda\varsigma s}
(\tilde{u}^\lambda_{\varsigma s}-\Delta_{\varsigma s})
\theta(\tilde{u}^\lambda_{\varsigma s}
-\Delta_{\varsigma s})/(4\pi\epsilon_\mathrm{r}\epsilon_0)\big]^{1/2}.
\end{equation}

At finite temperature, the full-$q$ polarization function is given by
\begin{align}
\tilde{\Pi}^{\varsigma s}_{\lambda,T}(\mathbf{q},\omega;
\tilde{\mu}^{\varsigma}_{\lambda})=
&\int_{-\infty}^\infty\frac{d\mu'\tilde{\Pi}^{\varsigma s}_{\lambda,T=0}(\mathbf{q},\omega)|_{\tilde{E}^\varsigma_{\lambda\mathrm{F}}
=\mu'}}{4k_BT\cosh^2[(\tilde{\mu}^{\varsigma}_\lambda-\mu')/2k_BT]}\nonumber\\
&\times\theta(|u'|-\lambda\varsigma s\gamma/2
-\Delta_{\varsigma s}),
\end{align}
where $\tilde{\mu}^{\varsigma}_{\lambda}$ is the chemical potential for
conduction or valence subband which can be obtained through Eq. (\ref{Cdensi}).

\end{document}